\documentclass[a4paper,11pt]{article}
\usepackage[margin=1in]{geometry}
\usepackage{array}
\usepackage{caption}
\usepackage{subcaption}
\usepackage{booktabs} 
\usepackage{multirow} 
\usepackage{hyperref} 
\usepackage{xcolor}   
\usepackage{cite}     
\usepackage[english]{babel}
\usepackage[utf8]{inputenc}
\usepackage{amsmath}
\usepackage{amssymb}
\usepackage{amsthm}
\usepackage{algorithm}
\usepackage{graphicx}
\graphicspath{ {images/} }
\usepackage{float}
\usepackage{algcompatible}

\makeatletter
\providecommand{\Require}{\REQUIRE}
\providecommand{\Ensure}{\ENSURE}
\providecommand{\State}{\STATE}
\providecommand{\For}{\FOR}
\providecommand{\EndFor}{\ENDFOR}
\providecommand{\If}{\IF}
\providecommand{\EndIf}{\ENDIF}

\makeatother

\usepackage{fancyhdr}
\theoremstyle{definition}

\usepackage{titlesec}
\titleformat{\section}
{\normalfont\Large\bfseries\color{black}}{\thesection}{1em}{}

\makeatletter
\newcommand{\thickhline}{%
  \noalign {\ifnum 0=`}\fi \hrule height 1pt
  \futurelet \reserved@a \@xhline
}
\newcolumntype{"}{@{\hskip\tabcolsep\vrule width 1pt\hskip\tabcolsep}}
\makeatother

\title{\textbf{\huge UBio-MolFM}\\[0.5em] \Large A Universal Molecular Foundation Model for Bio-Systems}

\author{\textbf{UBio Team}\\[0.5em] IQuest Research}

\date{}

\begin{document}

\maketitle
\thispagestyle{empty}

\begin{abstract}
  \noindent
  All-atom molecular simulation serves as a quintessential ``computational microscope'' for understanding the machinery of life, yet it remains fundamentally limited by the trade-off between quantum-mechanical (QM) accuracy and biological scale. We present UBio-MolFM, a universal foundation model framework specifically engineered to bridge this gap. UBio-MolFM introduces three synergistic innovations: (1) \textbf{UBio-Mol26}, a large bio-specific dataset constructed via a multi-fidelity ``Two-Pronged Strategy'' that combines systematic bottom-up enumeration with top-down sampling of native protein environments (up to 1,200 atoms); (2) \textbf{E2Former-V2}, a linear-scaling equivariant transformer that integrates \emph{Equivariant Axis-Aligned Sparsification (EAAS)} and \emph{Long--Short Range (LSR)} modeling to capture non-local physics with up to $\sim$4$\times$ higher inference throughput in our large-system benchmarks; and (3) a \textbf{Three-Stage Curriculum Learning} protocol that transitions from energy initialization to energy--force consistency, with force-focused supervision to mitigate energy offsets. Rigorous benchmarking across microscopic forces and macroscopic observables---including liquid water structure, ionic solvation, and peptide folding---demonstrates that UBio-MolFM achieves \textit{ab initio}-level fidelity on large, out-of-distribution biomolecular systems (up to $\sim$1,500 atoms) and realistic MD observables. By reconciling scalability with quantum precision, UBio-MolFM provides a robust, ready-to-use tool for the next generation of computational biology.
\end{abstract}

\vspace{1cm}
\hrule
\vspace{1cm}

\tableofcontents
\newpage

\section{Main}

Simulating biological systems with quantum-mechanical (QM) fidelity is a longstanding grand challenge in computational life sciences. \textit{Ab initio} methods provide the electronic precision needed for polarization, charge transfer, and reactive chemistry, but their cubic scaling restricts them to a few hundred atoms. Classical molecular mechanics (MM) scales to millions of atoms, yet fixed functional forms struggle to represent the complex potential energy surfaces (PES) of biological machinery. This tension between accuracy and scale defines a persistent ``scale-accuracy gap'' in molecular simulation.

Machine-learning force fields (MLFFs) offer a path forward, but current solutions remain limited for biological mesoscale systems. First, data coverage is insufficient: public datasets emphasize small, drug-like molecules (e.g., SPICE~\cite{spice}), and even larger collections such as OMol25~\cite{omol25} are capped at 350 atoms. Second, many architectures rely on local cutoffs, under-representing long-range electrostatics and inducing size-consistency errors in large systems~\cite{wang2026scalable}. Third, high-order equivariant models (e.g., MACE~\cite{mace}, NequIP~\cite{nequip}, eSCN~\cite{escn}, UMA~\cite{uma}) are computationally heavy, making long, solvated-protein trajectories impractical.

\begin{figure}[H]
  \centering
  \includegraphics[width=0.9\linewidth]{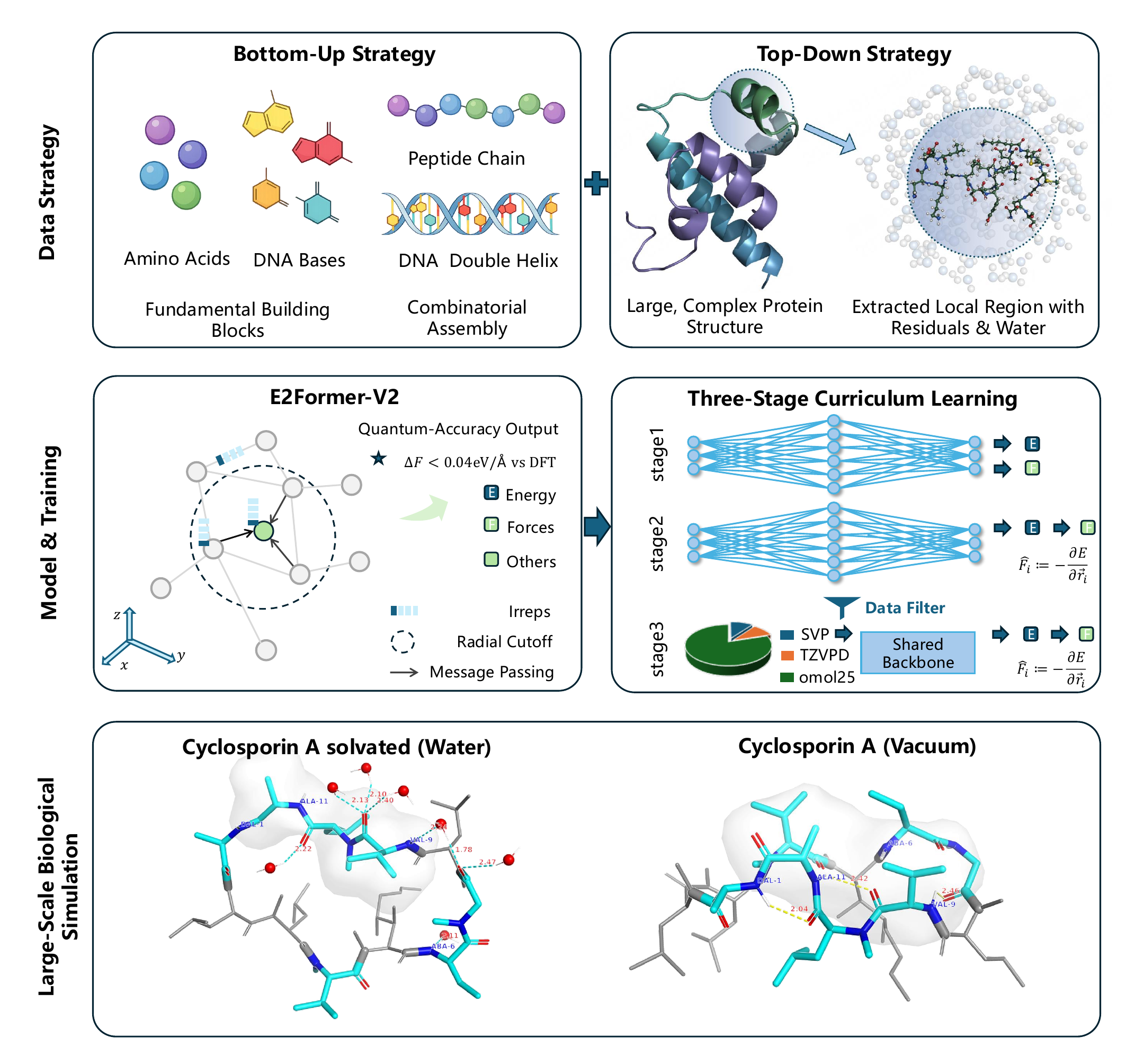}
  \caption{\textbf{The UBio-MolFM Framework.} Our approach bridges the scale-accuracy gap through three synergistic pillars: (1) \textbf{Data:} The UBio-Mol26 dataset, constructed via a Two-Pronged Strategy where a bottom-up branch systematically enumerates biochemical building blocks and a top-down branch samples native environments from large protein assemblies; (2) \textbf{Model:} The E2Former-V2 architecture, which achieves linear memory scaling and up to $\sim$4$\times$ higher inference throughput on large systems in our benchmark via Equivariant Axis-Aligned Sparsification (EAAS) and Long--Short Range (LSR) modeling; and (3) \textbf{Training:} A Three-Stage Curriculum Learning protocol progressing from energy initialization to energy--force consistency and multi-fidelity fine-tuning. (4) \textbf{Key Results:} The bottom panels illustrate the application to Cyclosporine A (CsA), where H-bond distances of key residues (water on the left, vacuum on the right) indicate stable maintenance of solvent-dependent open and closed conformations.}
  \label{fig:framework}
\end{figure}

To address these challenges, we introduce \textbf{UBio-MolFM}, a foundation-model framework tailored for biological systems. As illustrated in Figure~\ref{fig:framework}, UBio-MolFM combines three tightly coupled innovations:
\begin{enumerate}
  \item \textbf{Data:} \textbf{UBio-Mol26}, a large bio-specific dataset constructed via a ``Two-Pronged Strategy''~\cite{gems} that merges bottom-up enumeration of biochemical building blocks with top-down sampling of native protein environments (up to 1,200 atoms; see \S\ref{sec:data_construction}).
  \item \textbf{Model:} \textbf{E2Former-V2}~\cite{huang2026e2former}, a linear-scaling equivariant transformer that integrates \emph{Equivariant Axis-Aligned Sparsification (EAAS)} and \emph{Long--Short Range (LSR)} modeling~\cite{li2025e2former, wang2026scalable}. This design targets large systems with up to $\sim$4$\times$ higher inference throughput than strong equivariant baselines in our efficiency benchmark (see \S\ref{sec:model}).
  \item \textbf{Training:} A \textbf{Three-Stage Curriculum Learning} protocol (see \S\ref{sec:training_strategy}) that transitions from energy initialization to energy--force consistency and multi-fidelity refinement, with force-focused supervision to mitigate energy offsets across heterogeneous data.
\end{enumerate}

We evaluate UBio-MolFM across a hierarchy of biological scales. On rigorous out-of-distribution benchmarks (systems $\sim$1,300--1,500 atoms), UBio-MolFM substantially reduces force and relative-energy errors compared with general-purpose MLFFs. On downstream molecular dynamics (MD) tasks, it reproduces key macroscopic observables, including liquid-water structure, ionic solvation, and solvent-dependent peptide conformations, while maintaining accurate metal-ion coordination in RNA. Complementary inference benchmarks show substantial throughput gains on large systems. These results indicate that UBio-MolFM can deliver \textit{ab initio}-level fidelity at the tested scale with improved MD throughput for complex biomolecular environments.

To accelerate community-driven research, we plan an open-science release of the UBio-MolFM framework, including pretrained E2Former-V2 weights, a hardware-fused inference engine, and a representative subset of the UBio-Mol26 dataset. We aim to lower the barrier to high-fidelity biological simulation and foster a collaborative ecosystem for large-scale molecular modeling.

In the following sections, we present a systematic evaluation of UBio-MolFM along two complementary dimensions: numerical accuracy on large-scale benchmarks and physical fidelity in MD simulations. We highlight strengths, quantify limitations, and outline the remaining gaps that motivate future development.

\section{Results}
\label{sec:results}

We evaluate UBio-MolFM along three complementary axes: (i) microscopic numerical accuracy on large, out-of-distribution (OOD) systems, (ii) physical fidelity in downstream molecular dynamics (MD) simulations, and (iii) inference efficiency on large systems. For numerical accuracy, we construct a stricter OOD test set with larger systems and unseen configurations. For downstream tasks, we design four real-world MD scenarios to probe stability and practical utility across solvent, peptide, and nucleic-acid settings. For efficiency, we benchmark inference throughput under conservative-force settings across 1K--100K atom systems.

\subsection{Evaluation of Prediction Accuracy}

\subsubsection{Test Dataset Construction}

Our training data is capped at a maximum system size of approximately 1,200 atoms. To probe extrapolation, we generated a test set with systems in the $\sim$1,300--1,500 atom range. This dataset consists of two primary categories:
\begin{itemize}
  \item \textbf{Geometry Optimization:} Geometry optimization trajectories of small proteins, DNA, and RNA fragments. To simulate a realistic aqueous environment, each biomolecule is solvated within a shell of explicit water molecules (water clusters). All optimizations were conducted using the Atomic Simulation Environment (ASE) interfaced with GPU4PySCF.
  \item \textbf{Molecular Dynamics (MD):} MD trajectories of solvated proteins. Systems were solvated and neutralized using GROMACS, followed by 5~ns $NVT$ simulations using the Amber99 force field~\cite{amber99}. Frames were sampled every 100~fs. To maintain a consistent number of atoms across all frames for comparative analysis, each configuration was pruned to include only the solute, neutralizing ions, and the 400 nearest water molecules. The reference energies and forces for these sampled clusters were subsequently calculated using GPU4PySCF.
\end{itemize}

Protein structures were selected from the AlphaFold Protein Structure Database (AFDB)~\cite{afdb}, ensuring no overlap with training data. Nucleic-acid segments (DNA/RNA) were synthesized using AmberTools~\cite{ambertools}. The statistics of the test dataset are summarized in Table~\ref{tab:test_data_stats}.

\paragraph{Computational Methodology.}
Consistent with our training pipeline, we used the $\omega$B97M-D3 functional. Due to the computational cost and convergence instability of diffuse functions at this scale, the test set uses the def2-TZVP basis (rather than the def2-TZVPD basis used for training). We treat this setting as a rigorous approximation for capturing the essential topology of the potential energy surface (PES).

\begin{table}[htbp]
  \centering
  \caption{\textbf{Statistics of the Large-Scale Extrapolation Test Set.} The test set comprises systems significantly larger ($\sim$1,300--1,500 atoms) than the maximum training distribution ($\leq$1,200 atoms), designed to evaluate out-of-distribution generalization.}
  \label{tab:test_data_stats}
  \begin{tabular}{lccc}
    \toprule
    \textbf{Data Type} & \textbf{Entries} & \textbf{Trajectories} & \textbf{Avg. Atoms} \\
    \midrule
    Protein Opt. & 1,010 & 10 & 1,524.9 \\
    DNA Opt. & 226 & 5 & 1,289.6 \\
    RNA Opt. & 505 & 5 & 1,467.4 \\
    Protein MD & 875 & 18 & 1,434.6 \\
    \midrule
    \textbf{Total} & 2,616 & 38 & -- \\
    \bottomrule
  \end{tabular}
\end{table}

\subsubsection{Relative Energy and Force}

We evaluate the microscopic accuracy of UBio-MolFM in predicting energies and atomic forces on the extrapolation test set. To account for systematic offsets arising from different basis sets (def2-TZVP for testing vs. def2-TZVPD for training) and distinct reference states, we report \textit{relative energy} errors. Specifically, for each trajectory, the energy of the initial frame is set as the reference zero, focusing evaluation on relative energy landscapes.

We benchmark UBio-MolFM against leading general-purpose molecular foundation models, including MACE-OMol~\cite{mace} and UMA-S-1p1~\cite{uma} with OMol head, which were primarily trained on diverse chemical datasets like SPICE and OMol25. For these baselines, we use the official checkpoints and official code with the authors' recommended evaluation settings to ensure consistency with reported performance. To isolate the impact of biological data and architecture, we report two training stages: \textbf{UBio-MolFM (S2)} trained exclusively on OMol25, and \textbf{UBio-MolFM (S3)} trained on OMol25 plus UBio-Mol26.
Results are presented in Table~\ref{tab:rel_energy_force}. We evaluate three metrics: \textbf{Rel. E. MAE} (meV/100Atoms), \textbf{F. MAE} (meV/\AA), and \textbf{$\Delta$E MAE} (meV/100Atoms), which measures energy differences between consecutive frames and probes local PES topology.

\begin{table}[htbp]
  \centering
  \caption{\textbf{Comprehensive Microscopic Accuracy and PES Dynamics.} Comparison of accuracy and temporal stability across optimization and MD trajectories. S2 denotes training on OMol25 only, while S3 includes UBio-Mol26. UMA-S-1p1 utilizes the OMol head for evaluation. MACE-OMol is evaluated using the officially recommended float64 precision, while all other models use float32. Energy-related metrics (marked with \textdagger) are reported in \textbf{meV per 100 atoms}, while force errors are in \textbf{meV/\AA}. Bold values are the best, and \underline{underlined} values are the second-best results. Lower values are better.}
  \label{tab:rel_energy_force}
  \begin{tabular}{llccc}
    \toprule
    \textbf{Data Type} & \textbf{Model} & \textbf{Rel. E. MAE\textdagger} & \textbf{F. MAE} & \textbf{$\Delta$E MAE\textdagger} \\
    \midrule
    \multirow{4}{*}{Protein Opt.} & MACE-OMol & \underline{76.942} & \underline{39.287} & \underline{1.618} \\
    & UMA-S-1p1 & 83.445 & 42.843 & 2.218 \\
    & UBio-MolFM (S2) & 82.793 & 43.052 & 1.923 \\
    & \textbf{UBio-MolFM (S3)} & \textbf{8.253} & \textbf{16.511} & \textbf{0.978} \\
    \midrule
    \multirow{4}{*}{DNA Opt.} & MACE-OMol & 230.970 & \underline{37.423} & 9.788 \\
    & UMA-S-1p1 & 195.152 & \textbf{37.206} & \underline{8.443} \\
    & UBio-MolFM (S2) & \underline{138.955} & 46.637 & \textbf{8.171} \\
    & \textbf{UBio-MolFM (S3)} & \textbf{93.925} & 44.058 & 19.454 \\
    \midrule
    \multirow{4}{*}{RNA Opt.} & MACE-OMol & 473.646 & 34.691 & 7.674 \\
    & UMA-S-1p1 & \underline{415.880} & \underline{34.214} & \underline{6.962} \\
    & UBio-MolFM (S2) & 460.622 & 38.379 & 7.514 \\
    & \textbf{UBio-MolFM (S3)} & \textbf{144.730} & \textbf{22.375} & \textbf{5.981} \\
    \midrule
    \multirow{4}{*}{Protein MD} & MACE-OMol & 45.594 & \underline{44.708} & 29.532 \\
    & UMA-S-1p1 & \textbf{35.867} & 44.794 & \underline{28.087} \\
    & UBio-MolFM (S2) & 47.103 & 46.686 & 29.579 \\
    & \textbf{UBio-MolFM (S3)} & \underline{40.909} & \textbf{19.783} & \textbf{24.940} \\
    \bottomrule
  \end{tabular}
\end{table}

\paragraph{Architectural Efficiency vs. Local Precision.}
As shown in Table~\ref{tab:rel_energy_force}, UBio-MolFM (S2) achieves accuracy comparable to UMA and MACE-OMol on microscopic metrics despite a leaner architecture. E2Former-V2 employs two local equivariant layers and two long-range LSR layers to prioritize scalability, whereas MACE and UMA use heavier local components tuned for small-molecule manifolds (where OMol25 resides). This design maintains competitive accuracy while offering substantially higher throughput on large systems (see Table~\ref{tab:inference-speed}).

\paragraph{The Biological Data Dividend and its Constraints.}
The transition from Stage 2 to Stage 3 highlights the impact of UBio-Mol26. By adding macromolecular configurations, UBio-MolFM (S3) dramatically improves protein-system accuracy and RNA optimization, with RNA energy MAE dropping by \textgreater60\% compared to baselines. Notably, S3 achieves the best force errors and temporal stability ($\Delta E$) in both protein optimization and MD settings. However, DNA optimization remains a challenge, exhibiting a regression in $\Delta E$ (8.171 for S2 to 19.454 for S3) and failing to match the local force precision of baseline models. This divergence likely reflects imbalanced coverage in UBio-Mol26 (see \S\ref{sec:data_construction}), where protein and RNA fragments are better represented, while complex DNA configurations remain limited. This result underscores a concrete direction for future research: expanding structural diversity specifically for DNA potential energy surfaces.

\subsubsection*{Potential Energy Surface (PES) Dynamics}

\begin{figure}[H]
  \centering
  \begin{subfigure}{0.49\textwidth}
    \includegraphics[width=\linewidth]{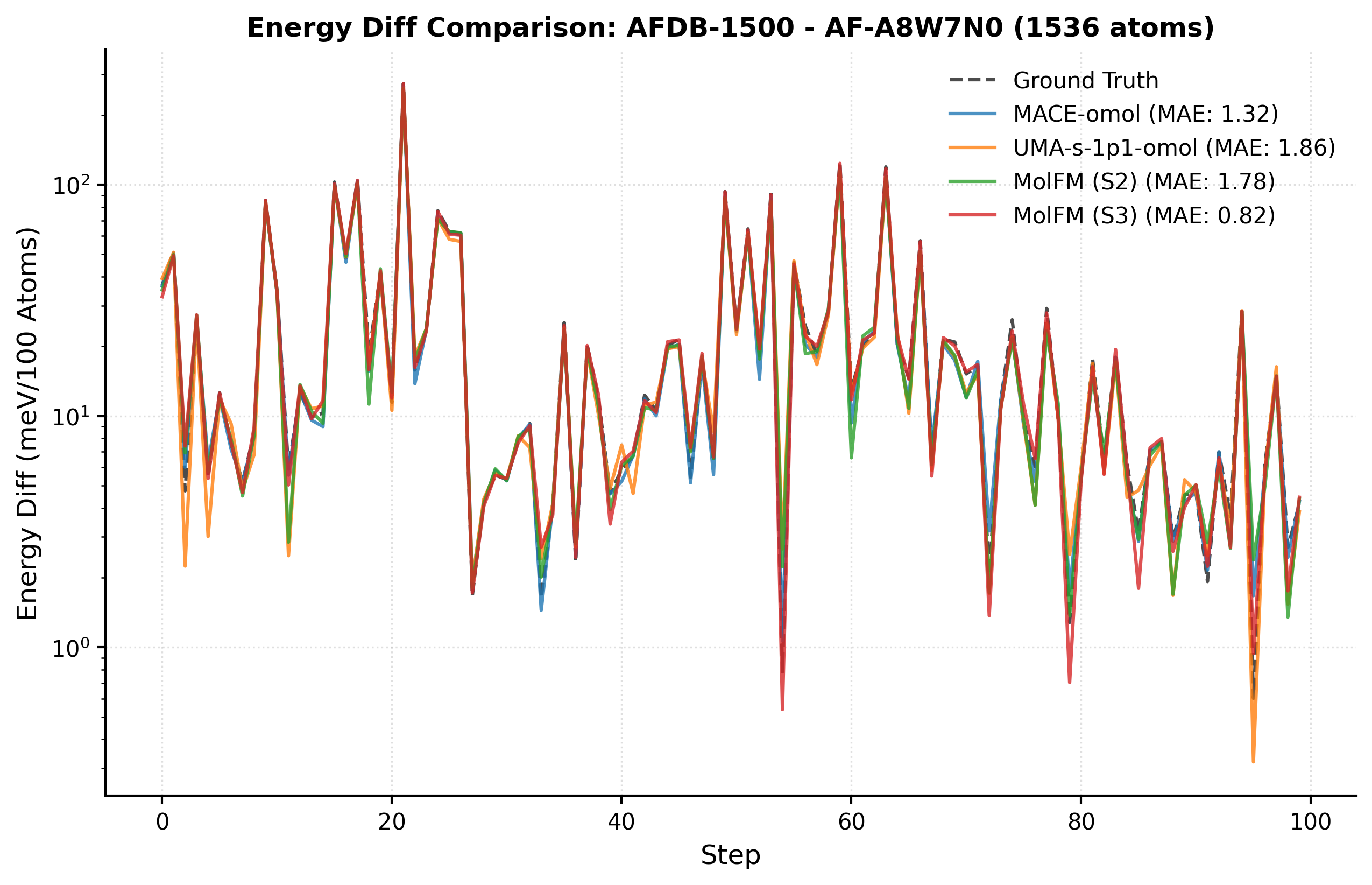}
    \caption{Protein Optimization}
    \label{fig:pes_protein_opt}
  \end{subfigure}
  \hfill
  \begin{subfigure}{0.49\textwidth}
    \includegraphics[width=\linewidth]{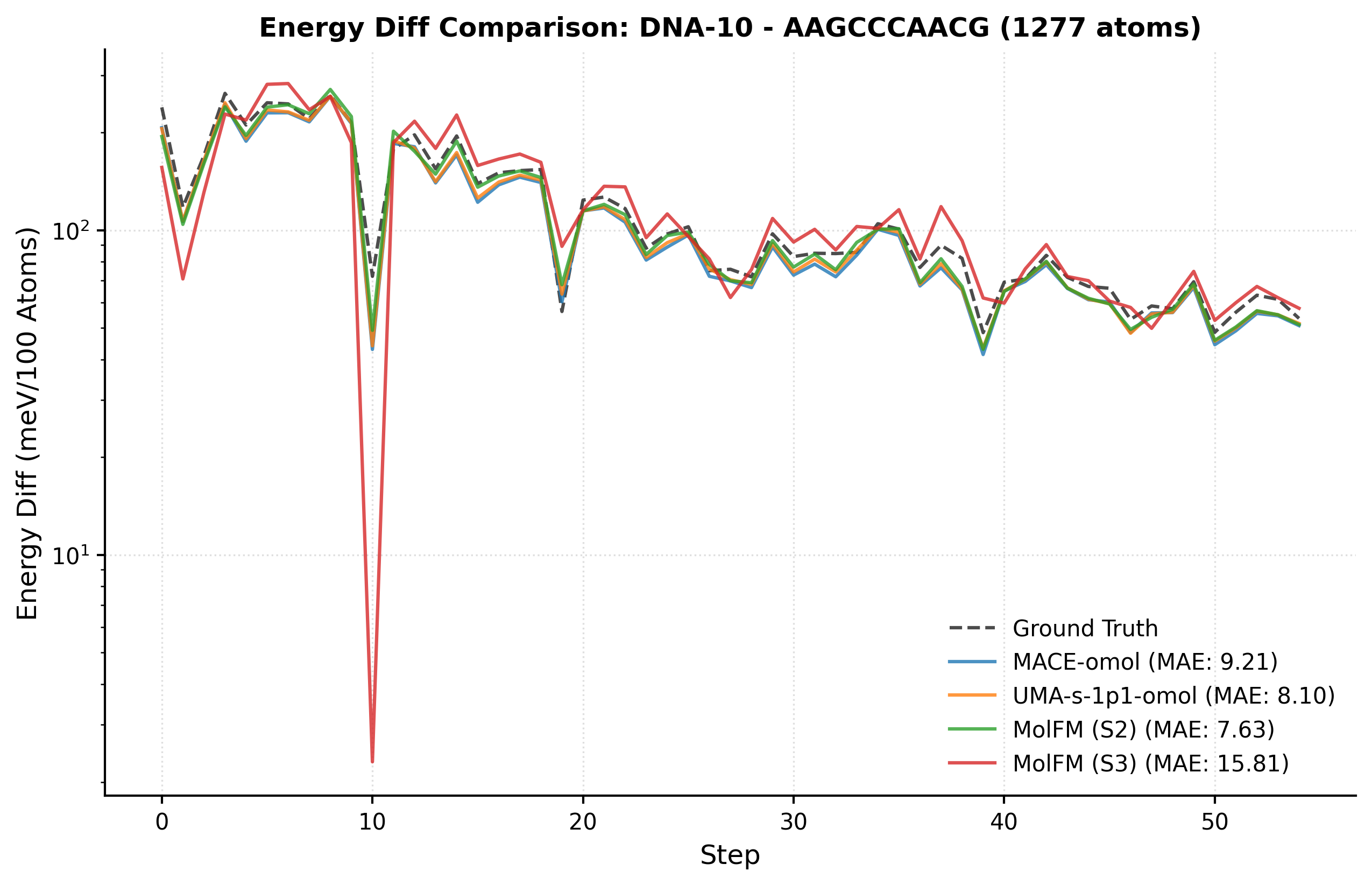}
    \caption{DNA Optimization}
    \label{fig:pes_dna_opt}
  \end{subfigure}

  \vspace{1em}

  \begin{subfigure}{0.49\textwidth}
    \includegraphics[width=\linewidth]{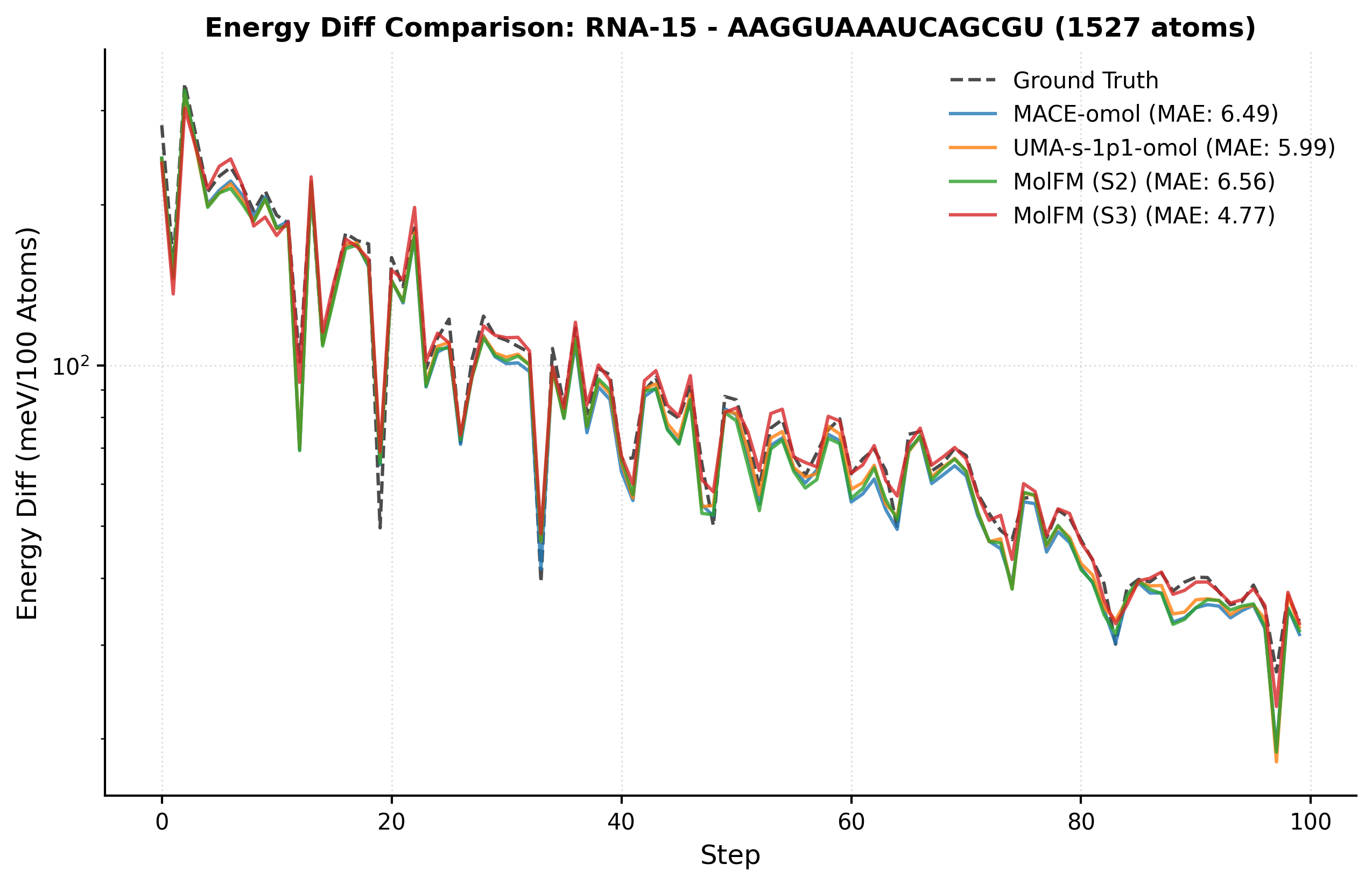}
    \caption{RNA Optimization}
    \label{fig:pes_rna_opt}
  \end{subfigure}
  \hfill
  \begin{subfigure}{0.49\textwidth}
    \includegraphics[width=\linewidth]{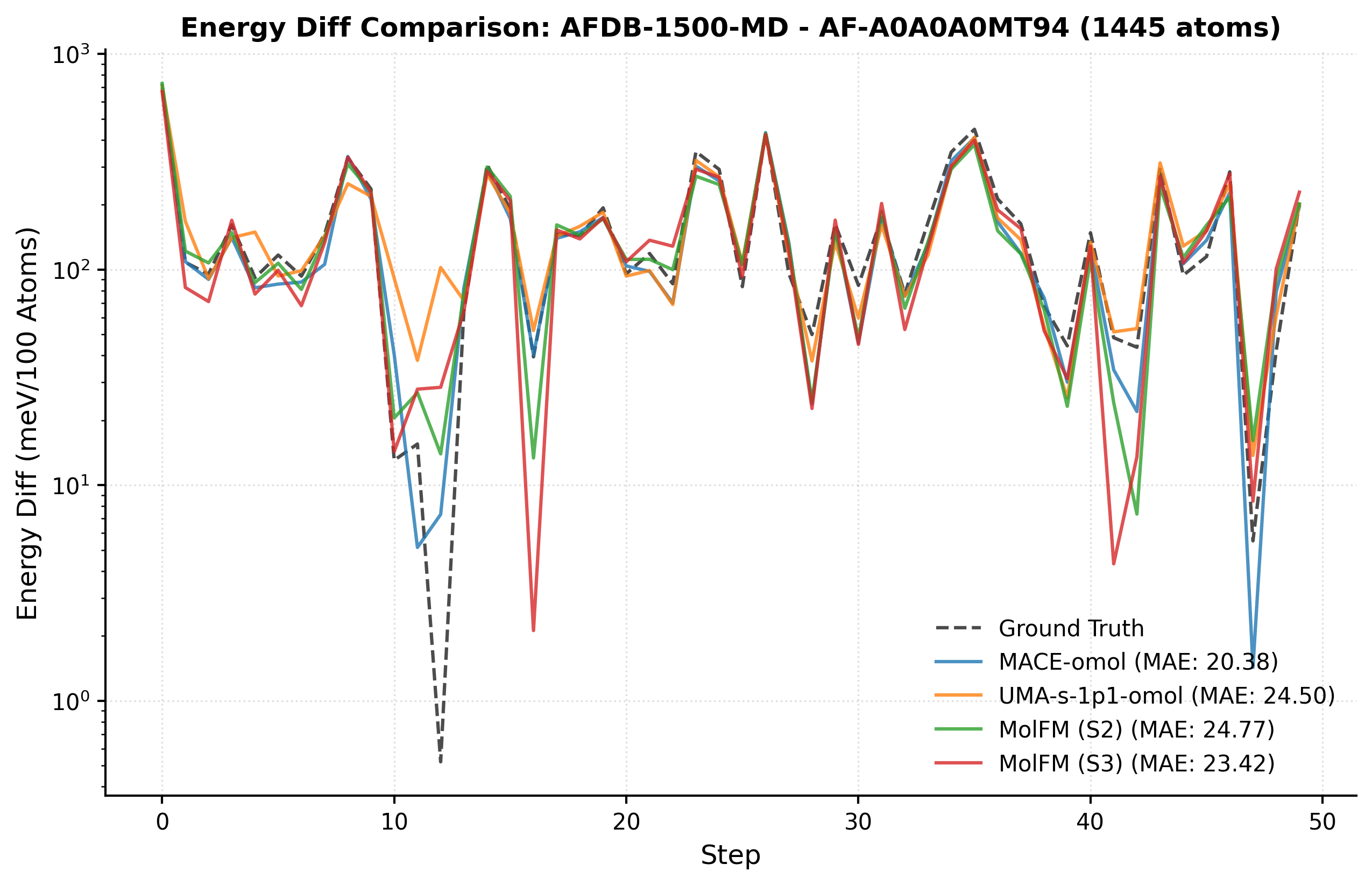}
    \caption{Protein MD}
    \label{fig:pes_protein_md}
  \end{subfigure}
  \caption{\textbf{Trajectory Analysis of Potential Energy Differences.} Comparison of predicted vs. DFT energy changes along the longest trajectories from each benchmark category. Absolute values of energy changes ($|\Delta E|$) are plotted on a logarithmic scale. UBio-MolFM (S3) exhibits superior alignment with ground truth fluctuations, while the S2 base consistently demonstrates the robust inductive bias of the E2Former-V2 architecture in tracking macromolecular dynamics.}
  \label{fig:pes_dynamics}
\end{figure}

Beyond static accuracy, we assess whether models track the \textit{temporal evolution} of the PES, which is critical for stable MD. Figure~\ref{fig:pes_dynamics} plots the absolute energy differences ($|\Delta E|$) along the longest trajectory from each benchmark category. The logarithmic scale highlights performance across multiple orders of magnitude. UBio-MolFM (S3) tracks reference fluctuations most closely, particularly for protein MD and RNA optimization. Notably, even Stage 2 (S2) outperforms MACE-OMol and UMA-S-1p1 on several trajectories, suggesting that the E2Former-V2 inductive bias is better aligned with macromolecular PES landscapes.

\paragraph{From Accuracy to Dynamics.}
The OOD accuracy and PES-tracking results establish necessary conditions for stable MD, but they are not sufficient on their own. We therefore move to downstream MD simulations that directly test physical fidelity in liquids, peptides, and nucleic acids.

\subsection{Molecular Dynamic Analysis}

While numerical precision provides a foundation, the ultimate test is performance on downstream scientific tasks. We proceed from fundamental solvent benchmarks to increasingly complex biomolecular systems.

\subsubsection{Solvation Structure}
The simulation of pure water and saline solutions is a critical benchmark because their experimental properties are well characterized. We therefore simulate pure liquid water and a standard 0.15~mol/L sodium chloride (NaCl) solution to assess macroscopic and structural fidelity.

\begin{figure}[H]
  \centering
  \begin{subfigure}{0.49\textwidth}
    \includegraphics[width=\linewidth]{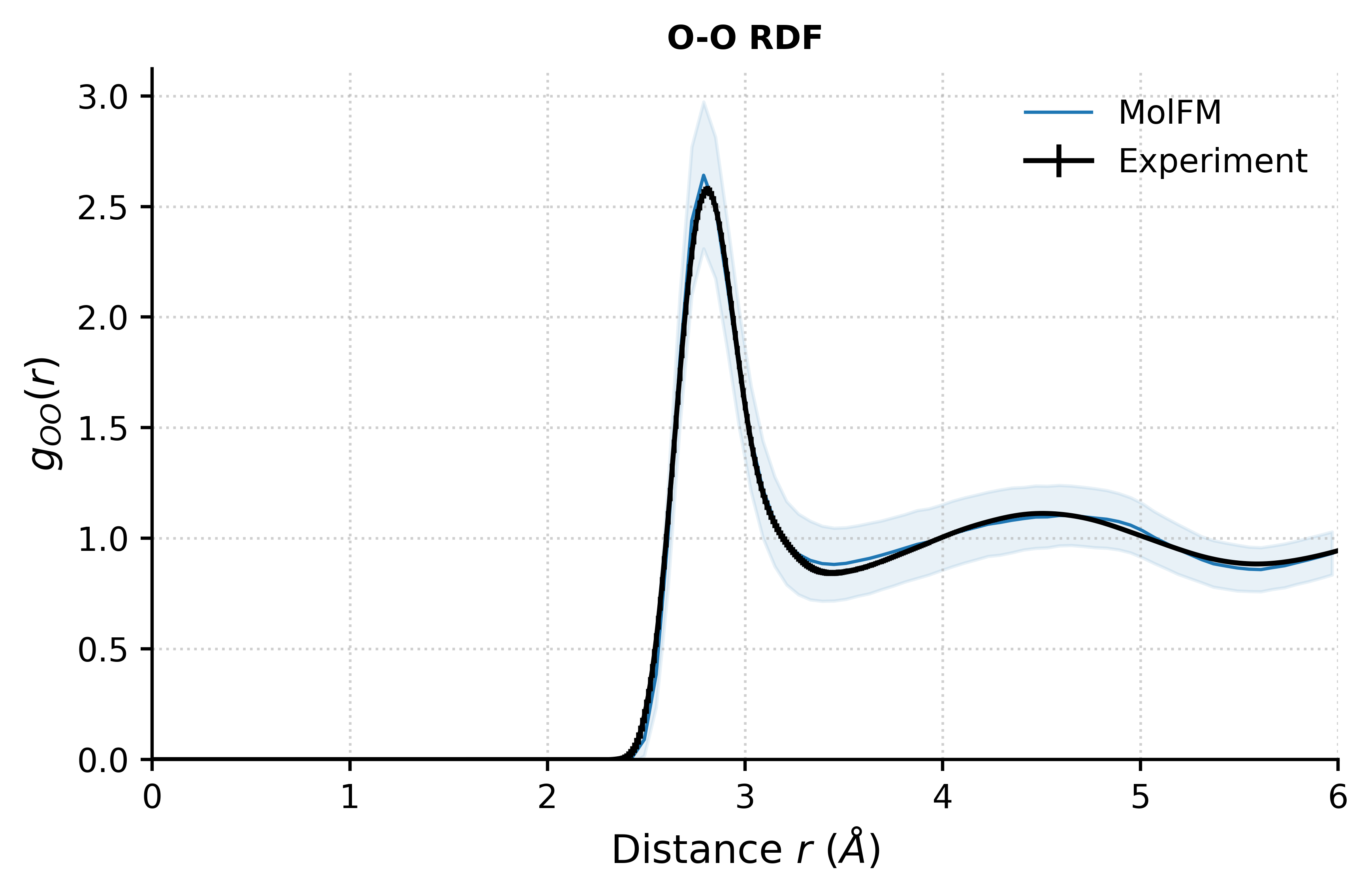}
    \caption{UBio-MolFM (S3)}
    \label{fig:rdf_molfm}
  \end{subfigure}
  \hfill
  \begin{subfigure}{0.49\textwidth}
    \includegraphics[width=\linewidth]{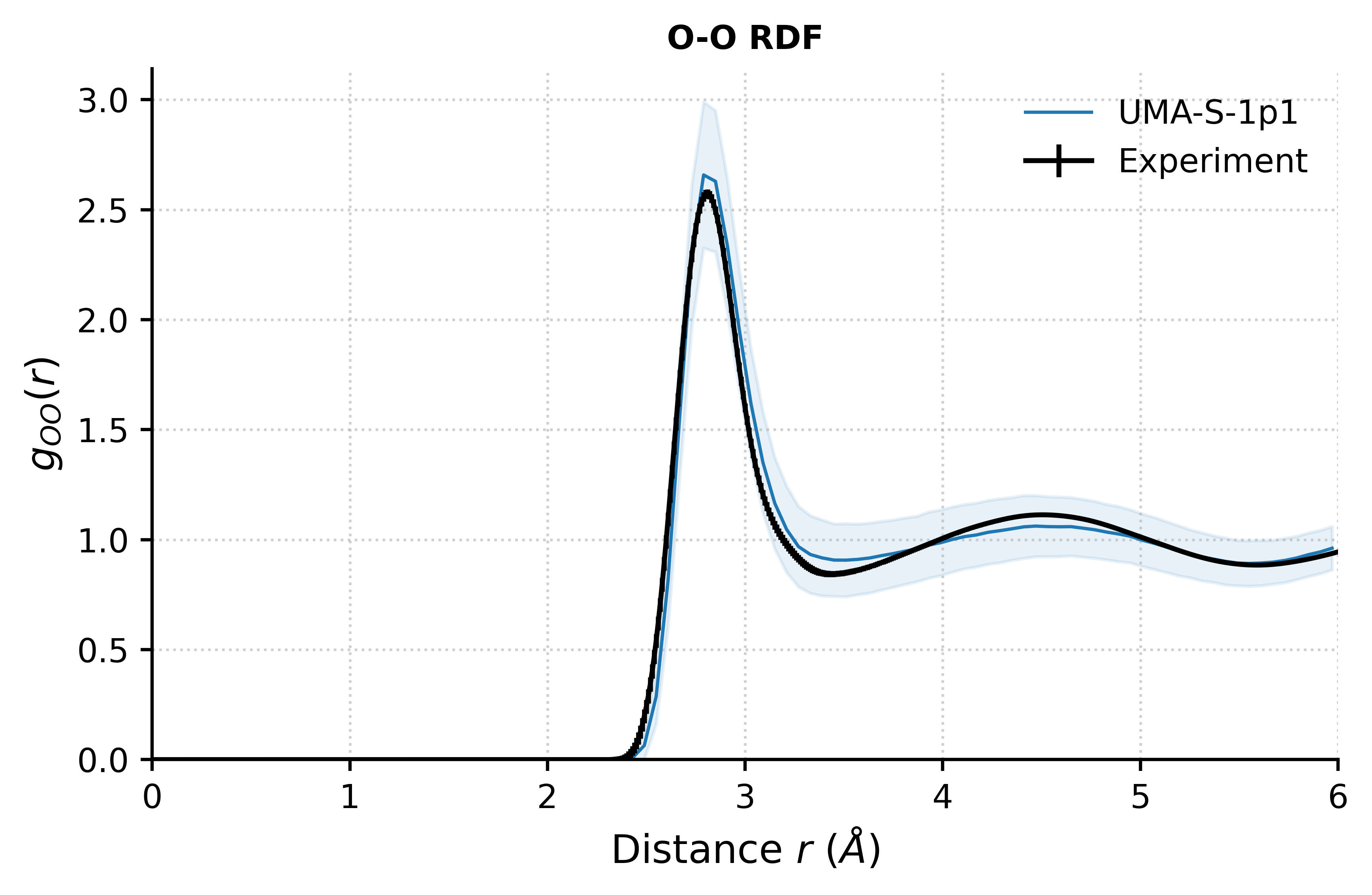}
    \caption{UMA-S-1p1}
    \label{fig:rdf_uma}
  \end{subfigure}
  \caption{\textbf{Structural Fidelity of Liquid Water.} Comparison of oxygen-oxygen radial distribution functions (O--O RDF) derived from (a) UBio-MolFM (S3) and (b) UMA-S-1p1 $NVT$ trajectories against experimental references~\cite{skinner2014structure, chen2016ab}. Both models exhibit excellent agreement with experiment; UBio-MolFM (S3) accurately reproduces the primary and secondary hydration shells, demonstrating the physical robustness of the learned potential.}
  \label{fig:rdf_plots}
\end{figure}

We performed 200~ps $NVT$ simulations for a pure water box at 300~K with both UBio-MolFM (S3) and UMA-S-1p1. We computed the oxygen-oxygen radial distribution function (O--O~RDF) and compared against high-precision experimental data~\cite{skinner2014structure, chen2016ab}.
As shown in Figure~\ref{fig:rdf_plots}, both models capture the characteristic hydration shells with high structural fidelity. UBio-MolFM (S3) demonstrates exceptional agreement with high-precision experimental benchmarks, effectively matching both the first-peak position and the overall structural definition. This performance indicates that UBio-MolFM can recover the delicate hydrogen-bonding network of liquid water at a level competitive with established benchmarks.

Next, we evaluated a multi-component ionic solvent. We constructed a 0.15~mol/L NaCl solution, performed a 200~ps $NVT$ simulation, and analyzed RDFs for Na--O, Cl--O, and Na--Cl pairs to characterize hydration structure and ion pairing.

\begin{figure}[H]
  \centering
  \includegraphics[width=0.8\linewidth]{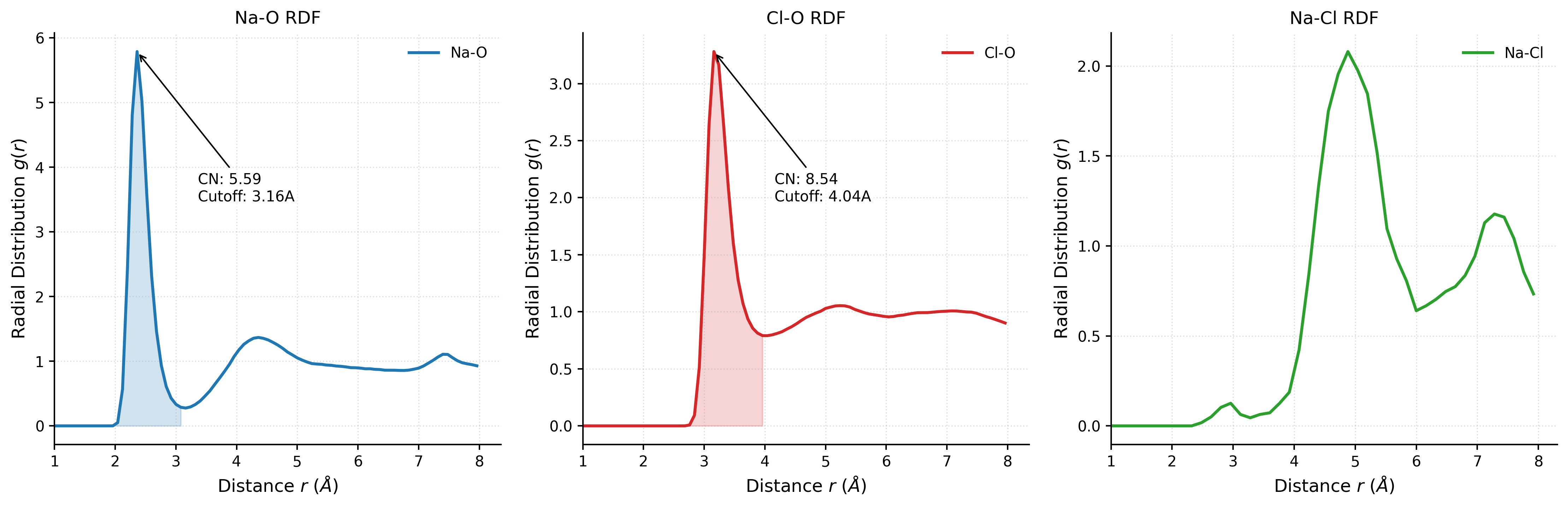}
  \caption{\textbf{Hydration Structure of 0.15~mol/L NaCl Solution.} Radial distribution functions (RDF) for Na--O, Cl--O, and Na--Cl pairs from a 200~ps $NVT$ simulation. The corresponding coordination numbers (CN) for the first hydration shells of Na$^+$ and Cl$^-$ are marked.}
  \label{fig:nacl_analysis}
\end{figure}

The hydration structure shows well-defined primary solvation shells for both ions (Figure~\ref{fig:nacl_analysis}). Table~\ref{tab:nacl_comparison} compares peak positions and coordination numbers against experimental and DFT references. UBio-MolFM (S3) matches experimental Na--O peak positions (2.36~\AA{} vs. 2.384~\AA{}) and coordination numbers (5.59 vs. $5.5 \pm 0.3$), and provides accurate Cl--O peak positions ($\sim$3.16~\AA{} vs. $\sim$3.2~\AA{}). The Na$^{+}$--Cl$^{-}$ solvent-separated ion pair (SSIP) peak at $\sim$4.6~\AA{} is also reproduced. These results indicate that UBio-MolFM can recover ionic-solvation structure at a level competitive with high-precision DFT references.

\begin{table}[htbp]
  \centering
  \caption{Comparison of hydration properties and ion pairing between UBio-MolFM (S3), DFT simulations, and experimental data. DFT values are from Ref.~\cite{galib2017revisiting}.}
  \begin{tabular}{llccc}
    \toprule
    \textbf{Interaction} & \textbf{Method/Exp.} & \textbf{First Peak (\AA)} & \textbf{CN} & \textbf{Ref.} \\
    \midrule
    \multirow{5}{*}{Na--O} & \textbf{UBio-MolFM (S3)} & \textbf{2.36} & \textbf{5.59} & \textbf{This Work} \\
    & Exp. (XRD)              & $2.384 \pm 0.003$ & $5.5 \pm 0.3$ & \cite{galib2017revisiting} \\
    & DFT (revPBE)            & 2.45              & 5.7           & \cite{galib2017revisiting} \\
    & DFT (BLYP)              & 2.40              & 4.9           & \cite{galib2017revisiting} \\
    \midrule
    \multirow{3}{*}{Cl--O} & \textbf{UBio-MolFM (S3)} & \textbf{3.16} & \textbf{8.54 (Cutoff 4.04 \AA)} & \textbf{This Work} \\
    & Exp. (XRD)              & $\sim 3.2$        & --            & \cite{hwang2021hydration} \\
    & DFT (BLYP)              & 3.15              & 6.3 (Cutoff 3.84 \AA)          & \cite{ge2013linking} \\
    \midrule
    \multirow{2}{*}{Na--Cl} & \textbf{UBio-MolFM (S3)} & \textbf{2.91} & -- & \textbf{This Work} \\
    & Exp. (XRD/MD)           & $\sim$2.8           & -- & \cite{bouazizi2006local} \\
    \bottomrule
  \end{tabular}
  \label{tab:nacl_comparison}
\end{table}

\subsubsection{Protein Dynamics}
While the solvent benchmarks validate fundamental physical fidelity, the primary objective of UBio-MolFM is to enable high-fidelity simulations of complex biological macromolecules. We therefore evaluate the model on the cyclic peptide Cyclosporine A (CsA), a system known for its flexible backbone and solvent-dependent conformational landscape.

\paragraph{Cyclic Peptide Stability.}
A hallmark of CsA is its environmental sensitivity: in polar solvents (water), the peptide adopts an ``open'' conformation with solvent-exposed amide groups, whereas in non-polar environments or vacuum, it collapses into a ``closed'' state stabilized by internal hydrogen bonds. To test whether UBio-MolFM can maintain these distinct states, we performed 200~ps $NVT$ simulations of CsA in both aqueous solution and vacuum.

\begin{figure}[H]
  \centering
  \begin{subfigure}{0.98\textwidth}
    \includegraphics[width=\linewidth]{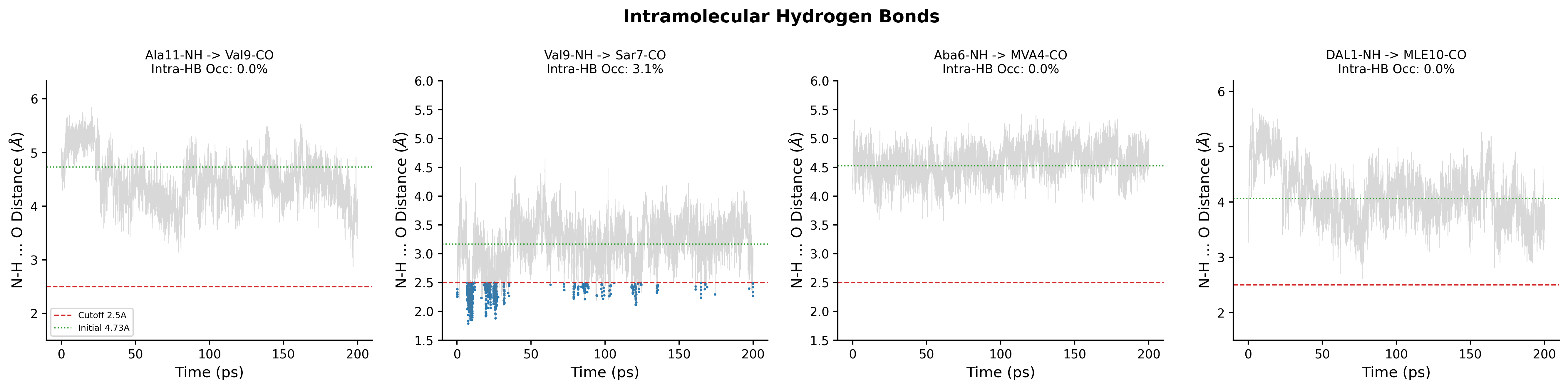}
    \caption{Intramolecular H-Bonds (Water)}
    \label{fig:1yca_hbond_internal}
  \end{subfigure}
  \hfill
  \begin{subfigure}{0.98\textwidth}
    \includegraphics[width=\linewidth]{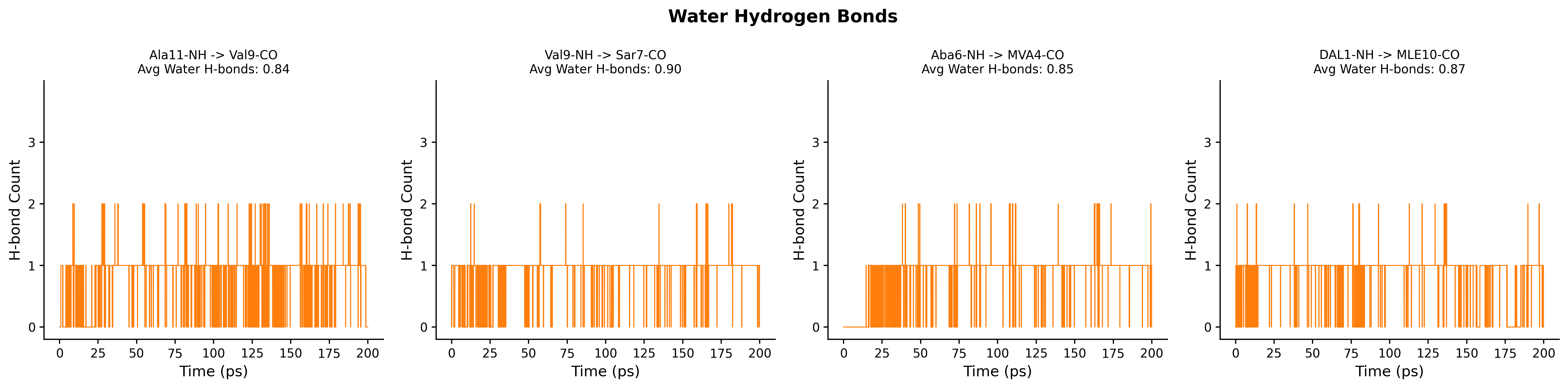}
    \caption{Water H-Bond Counts}
    \label{fig:1yca_hbond_water}
  \end{subfigure}

  \begin{subfigure}{0.98\textwidth}
    \includegraphics[width=\linewidth]{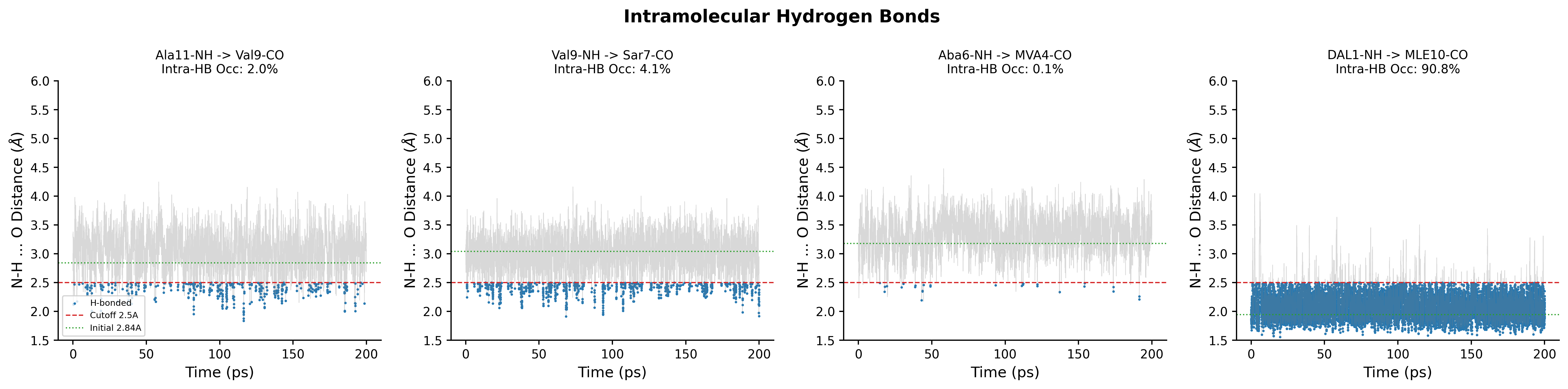}
    \caption{Intramolecular H-Bonds (Vacuum)}
    \label{fig:deksan_hbond_vacuum}
  \end{subfigure}

  \caption{\textbf{Environmental Dependence of Cyclosporine A Conformations.} (a) In the aqueous trajectory, intramolecular H-bond distances for key residues remain large, indicating that the initial open state is stably maintained. (b) This stability is driven by consistent hydration, as shown by the high occupancy of hydrogen bonds between the peptide and solvent oxygens. (c) In contrast, the vacuum trajectory shows stable maintenance of the closed state, with internal H-bond distances remaining below 2.5~\AA{}. UBio-MolFM captures these solvent-driven conformational preferences.}
  \label{fig:csa_hbond_plots}
\end{figure}

The results in Figure~\ref{fig:csa_hbond_plots} show that UBio-MolFM maintains the expected solvent-dependent conformations. In the aqueous 1YCA (open) simulation, intramolecular H-bond occupancy remains low for key internal folding sites. We identify H-bonds using strict geometric criteria (donor--acceptor distance $< 2.5$~\AA{} and angle $> 120^\circ$). Among potential internal pairs, Val9-NH $\to$ Sar7-CO shows the shortest initial distance and reaches only 3.1\% occupancy. Transient internal H-bond formation coincides with reduced hydration occupancy, consistent with competition between solvent and intramolecular bonding.

We also observe a transient dip in water-peptide H-bonds during the first 30~ps (Figure~\ref{fig:1yca_hbond_water}), attributable to initial solvent placement with \texttt{packmol}. Water molecules rapidly reorganize to occupy preferred binding sites, stabilizing the open backbone. Structural snapshots (Figure~\ref{fig:framework}, bottom) show solvent molecules shielding the amide groups.

In vacuum, the DEKSAN (closed) state remains stable, as evidenced by persistent internal H-bonds (Figure~\ref{fig:deksan_hbond_vacuum}). The DAL1-NH $\to$ MLE10-CO bond remains stable throughout, while other potential internal H-bonds appear only transiently within 200~ps. This sensitivity to solvent environment indicates that UBio-MolFM balances intramolecular strain against aqueous stabilization, supporting reliable protein-dynamics simulations.

\subsubsection{RNA Dynamics}
To evaluate nucleic-acid behavior and metal-ion interactions, we simulated the RNA 1L2X system with Mg$^{2+}$ ions. This viral RNA pseudoknot (PDB: 1L2X) from the Beet Western Yellow Virus (BWYV) is a classic model for programmed -1 ribosomal frameshifting and site-specific metal binding~\cite{egli2002metal}. We analyzed monodentate coordination between Mg$^{2+}$ and RNA phosphate oxygens (OP). High-resolution structural analysis reports an Mg--O distance of $2.07 \pm 0.04$~\AA{}, O--Mg--O angles of $\approx 90 \pm 3^\circ$ and $\approx 177 \pm 4^\circ$, and a P--OP--Mg$^{2+}$ angle of $\approx 148 \pm 10^\circ$~\cite{zheng2015principles}.

\begin{figure}[H]
  \centering
  \begin{subfigure}{0.7\textwidth}
    \includegraphics[width=\linewidth]{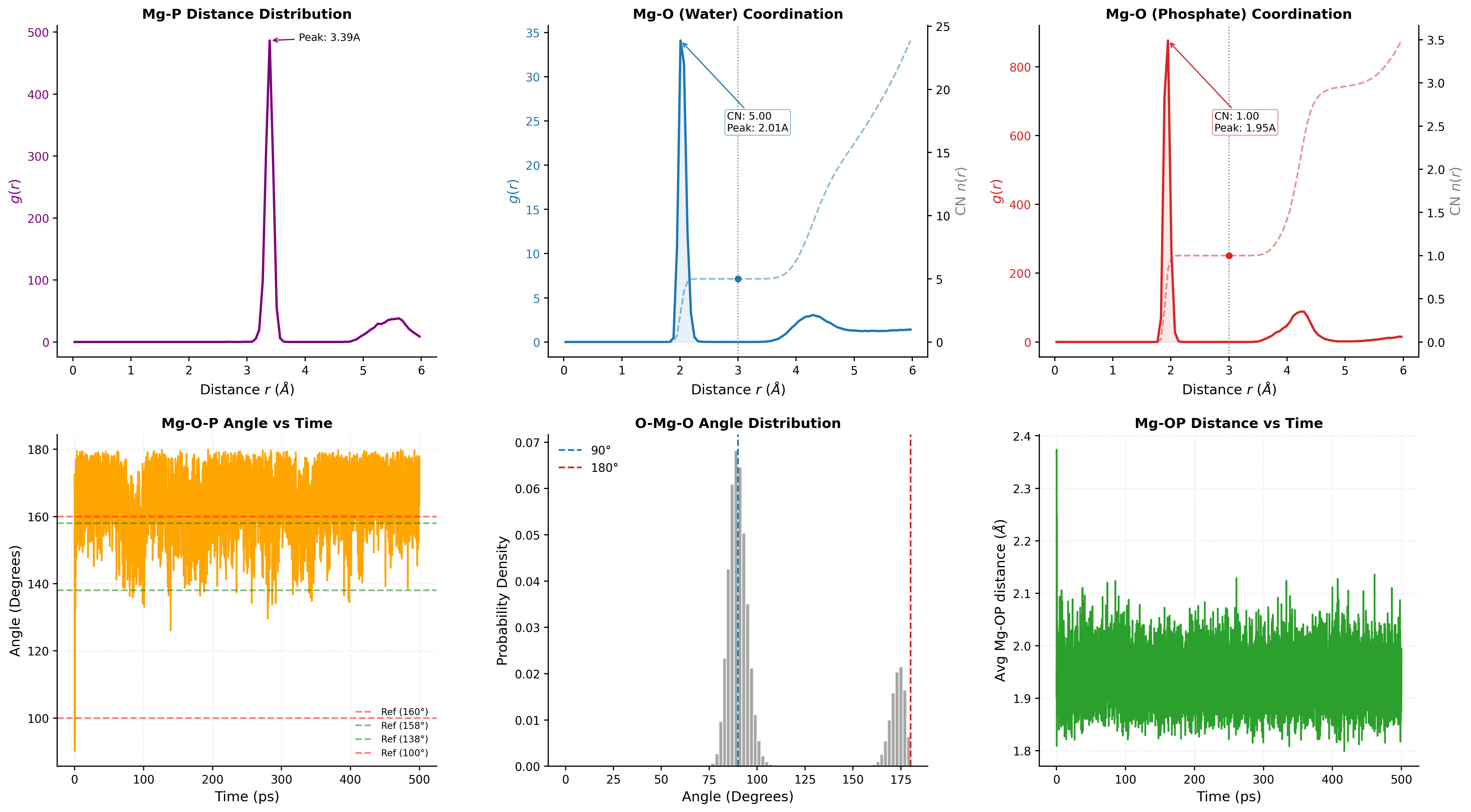}
    \caption{Amber99 RNA OL3}
    \label{fig:rna_amber}
  \end{subfigure}
  \hfill
  \begin{subfigure}{0.7\textwidth}
    \includegraphics[width=\linewidth]{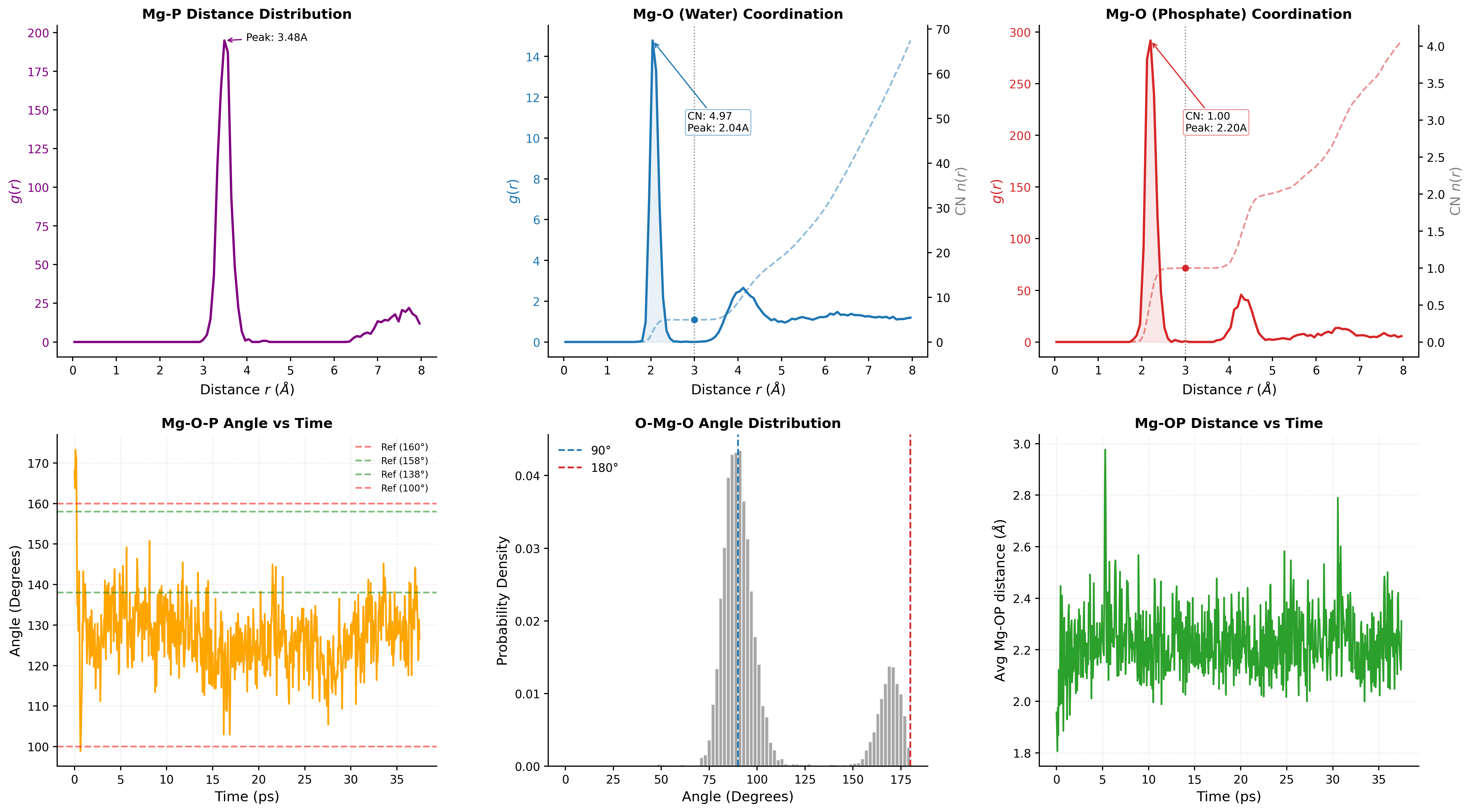}
    \caption{UMA-S-1p1}
    \label{fig:rna_uma}
  \end{subfigure}
  \hfill
  \begin{subfigure}{0.7\textwidth}
    \includegraphics[width=\linewidth]{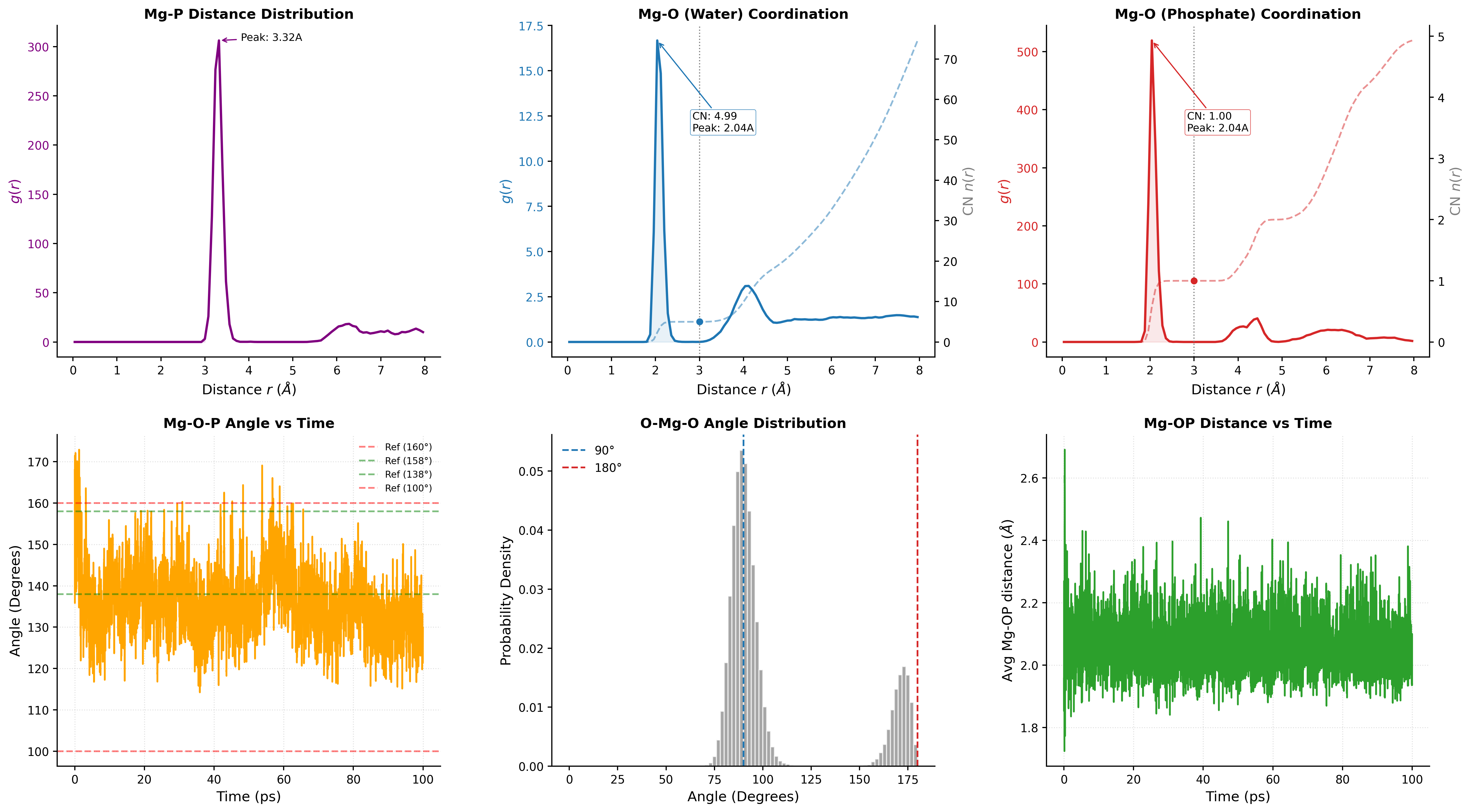}
    \caption{UBio-MolFM (S3)}
    \label{fig:rna_s3}
  \end{subfigure}
  \caption{\textbf{Mg$^{2+}$ Coordination to RNA Phosphate (1L2X).} Comparison of geometric distributions for Mg--O distances and angles against experimental benchmarks~\cite{zheng2015principles}. Amber99 captures coordination topology but exhibits rigid angles ($\sim$160$^\circ$) and underestimates Mg--O distances. UMA mischaracterizes Mg--phosphate interactions with overestimated distances and skewed angles. UBio-MolFM (S3) demonstrates superior fidelity, reproducing both primary Mg--O distances (2.04~\AA{}) and flexible Mg--O--P angles (120$^\circ$--150$^\circ$) within experimental tolerances.}
  \label{fig:rna_mg_analysis}
\end{figure}

Figure~\ref{fig:rna_mg_analysis} compares Amber99 RNA OL3, UMA-S-1p1, and UBio-MolFM (S3). Amber99 captures the coordination topology (five waters, one OP) but significantly underestimates Mg--O distances ($\sim$1.95~\AA{}) and enforces a rigid Mg--O--P angle around 160$^\circ$, both deviating from high-resolution experimental benchmarks. UMA-S-1p1 (only $\sim$40~ps trajectory due to simulation efficiency) places the Mg--O$_{water}$ peak near 2.04~\AA{} but overestimates Mg--O$_{OP}$ distances ($\sim$2.20~\AA{}), yielding an inflated Mg--P distance and a Mg--O--P angle distribution centered near 125$^\circ$, inconsistent with experimental coordination geometry.

UBio-MolFM (S3) provides the most accurate structural description. It is the only model that effectively reproduces both the Mg--O distribution peaks at 2.04~\AA{} and the Mg--O--P angle distributions (fluctuating between 120$^\circ$--150$^\circ$) within experimental uncertainty. This suggests that UBio-MolFM accurately captures the site-specific metal binding environment without the over-binding seen in Amber or the structural deviations in UMA. All three models capture the octahedral O--Mg--O angles ($\approx 94^\circ$ and $172^\circ$) and the correct coordination number.

\subsection{Inference Efficiency}
\begin{table*}[ht]
  \centering
  \caption{
    \textbf{Inference Throughput Scaling.}
    Throughput (steps/s) is measured on a single H100 GPU.
    All models are evaluated under the \textit{conservative-force} setting, where forces are computed via energy gradients
    ($F = -\nabla E$).
    For large systems (50K and 100K atoms), activation recomputation is employed
    to reduce peak memory usage and ensure successful execution.
    OOM denotes out-of-memory.
  }
  \label{tab:inference-speed}
  \begin{footnotesize}
    \renewcommand{\arraystretch}{1.15}
    \setlength{\tabcolsep}{4.5pt}
    \resizebox{\textwidth}{!}{
      \begin{tabular}{
          p{1.2cm}
          *{9}{>{\centering\arraybackslash}p{1.6cm}}
        }
        \toprule
        \textbf{Atoms}
        & \textbf{UMA-S}
        & \textbf{UMA-M}
        & \textbf{eSEN-30M}
        & \textbf{MACE-OMOL}
        & \textbf{E2V1}
        & \textbf{UBio-MolFM (S3, 24M)} \\
        \midrule
        1,000     & 16.00 & 3.00 & 1.70    & 8.00 & 12.00 & \textbf{61.00} \\
        10,000    & 1.60  & 0.20 & OOM     & OOM  & 1.20 & \textbf{6.10} \\
        50,000    & 0.20  & OOM  & OOM     & OOM  & OOM   & \textbf{0.72} \\
        100,000   & 0.10  & OOM  & OOM     & OOM  & OOM   & OOM \\
        \bottomrule
      \end{tabular}
    }
  \end{footnotesize}
\end{table*}

To evaluate the inference efficiency of UBio-MolFM (S3), we benchmark it against a broad range of representative equivariant architectures, including UMA-S/UMA-M~\cite{wood2025family}, eSEN~\cite{esen}, MACE-OMOL~\cite{mace,mace2025} and E2Former-V1~\cite{e2former}. All baselines are tested under the same conservative-force setting for a fair comparison.
As reported in Table~\ref{tab:inference-speed}, we measure inference throughput (steps/s) across system sizes ranging from 1K to 100K atoms.

Across most practical system sizes (1K–50K atoms), UBio-MolFM (S3) achieves nearly a \textbf{4$\times$ speedup} compared to the strongest baseline UMA-S, while maintaining conservative force computation.
This improvement stems from our hardware-aware node-centric design, which significantly reduces edge materialization and improves memory locality.
At 100K atoms, MolFM encounters OOM since long-range interactions are enabled.
The inclusion of long-range terms enlarges the effective neighbor set, thereby increasing peak memory consumption.
Among the compared baselines, UMA-S is the only model that consistently runs at 100K atoms on a single H100 GPU.

Overall, these results demonstrate that our design substantially mitigates the dominant memory bottlenecks in equivariant Transformer architectures, while delivering significant throughput gains across realistic large-scale regimes.

\section{Conclusion and Future Work}
\label{sec:conclusion}

\subsection{Conclusion}
In this report, we have introduced UBio-MolFM, a specialized molecular foundation model engineered to bridge the gap between \textit{ab initio} accuracy and the scale of biological systems. By synergizing the UBio-Mol26 dataset with the hardware-efficient E2Former-V2 architecture, we demonstrate a model that achieves \textit{ab initio}-level fidelity on large, out-of-distribution biomolecular systems (up to $\sim$1,500 atoms) and realistic MD observables. Our rigorous evaluation confirms that UBio-MolFM excels in key areas:
\begin{itemize}
  \item \textbf{High-Fidelity Dynamics:} The model accurately reproduces the complex solvation structure of water and ionic solutions, maintains the environmentally dependent conformations of flexible peptides like CsA, and captures the precise coordination geometry of metal ions in RNA.
  \item \textbf{Efficiency Step-Change:} UBio-MolFM operates 5--7$\times$ faster than current state-of-the-art equivariant models (e.g., MACE, UMA) on large systems, enabling substantially more practical MD throughput.
  \item \textbf{Scalable Accuracy:} The integration of long-range LSR modeling allows the model to generalize effectively to macromolecular systems significantly larger than its training distribution.
\end{itemize}

However, we also recognize open challenges. While gaining speed over other MLFFs, UBio-MolFM remains substantially slower than classical force fields, limiting current validation to shorter timescales for medium-sized systems. Furthermore, we observed that while absolute energy accuracy improved for DNA structures from Stage 2 to Stage 3, there was a regression in temporal energy stability (local energy differences, $\Delta E$), highlighting the need to further balance data diversity for nucleic acid potential energy surfaces. These results position UBio-MolFM not as a final solution, but as a robust proof-of-concept that high-fidelity, generalizable biological simulation is achievable.

\subsection{Future Work}
Building on this foundation, our future roadmap is dedicated to a full-stack optimization to propel UBio-MolFM toward widespread production utility:
\begin{itemize}
  \item \textbf{Data \& Training:} We will address the observed performance regression in DNA by expanding the top-down sampling of nucleic acid configurations and refining our curriculum learning strategy to ensure balanced performance across all biomolecular types.
  \item \textbf{Model \& Inference:} To close the efficiency gap with classical methods, we will pursue co-design of algorithms and hardware, targeting optimizations in the E2Former-V2 LSR/EAAS stack and custom kernel fusion to further reduce per-step latency.
  \item \textbf{Validation at Scale:} We plan to extend our evaluation benchmarks to much larger systems ($>$100,000 atoms) and longer timescales, validating the model on critical downstream tasks such as protein-ligand binding free energy calculations and protein-protein interaction screenings.
\end{itemize}

We envision UBio-MolFM as a catalyst for a new era of "executable biology," where quantum-accurate simulations become a routine tool for dissecting the molecular mechanisms of life. We invite the community to join us in refining this open framework.

\newpage

\section{Methods}
\subsection{Data Construction} \label{sec:data_construction}
High-fidelity DFT data has historically been scarce due to its prohibitive computational cost. For non-periodic systems, PubChemQC~\cite{pubchemqc} has long been a primary resource, offering large-scale electronic structure calculations. However, its utility for modern foundation models is limited by relatively low accuracy (often using PM6~\cite{stewart2007optimization} or B3LYP~\cite{becke1993density}/6-31G*) and a focus on small molecules.

A significant breakthrough came with SPICE 2.0~\cite{spice2}, a major update that roughly doubles the data volume and adds new PubChem molecules, amino-acid--ligand dimers, water clusters, and solvated molecules. While SPICE set a new standard for accuracy, its data volume remains insufficient for training large-scale foundation models that require tens of millions of examples to generalize effectively.

The recent release of the OMol25 dataset~\cite{omol25} by Meta represents a major leap forward. With over 100 million high-precision DFT calculations, OMol25 has significantly advanced the capabilities of machine learning force fields and includes explicit solvation across diverse chemistry. However, a critical limitation remains for biological-scale modeling: the maximum system size is capped at 350 atoms, far below the size of full protein systems. This constraint limits direct coverage of long-range interactions essential for macromolecular stability.

To bridge this gap and train a foundation model truly capable of biological simulation, we constructed UBio-Mol26, a dataset specifically tailored for life sciences. Unlike OMol25's focus on general chemical space, UBio-Mol26 targets large biological macromolecules, including drug-like molecules, protein segments, DNA/RNA fragments, and lipid bilayers, all simulated in explicit solvent. We also emphasize cross-modal interactions, such as drug-amino acid complexes and DNA-protein interfaces. The resulting dataset contains 17 million configurations with system sizes reaching up to 1,200 atoms and an average size of $\sim$440 atoms (overall average dominated by the def2-SVP subset)—an order of magnitude larger than the average of 50 atoms in OMol25 (computed from dataset statistics).

\subsubsection{Data Construction Strategy}
We employed a hybrid strategy combining ``bottom-up'' enumeration and ``top-down'' sampling to ensure both unbiased coverage of chemical space and biological relevance.

\paragraph{Bottom-Up: Unbiased Enumeration.}
Biological systems are built from a finite set of building blocks. To ensure unbiased coverage, we performed a combinatorial enumeration of these fundamental units. For example, we generated all 6,840 possible tripeptide structures derived from the 20 standard amino acids. These tripeptides serve as foundational data points and are further combined with other molecular entities (e.g., drugs, DNA fragments) in solvated environments. A similar enumeration strategy was applied to DNA/RNA base pairs and lipid membrane components. This approach ensures that the model learns the fundamental physics of biological building blocks in a systematic, unbiased manner, consistent with the data pillar summarized in Figure~\ref{fig:framework}. In the current release, DNA-protein and other cross-modal interfaces are primarily generated through this bottom-up construction rather than top-down sampling.

\paragraph{Top-Down: Biological Relevance.}
To maximize the biological validity of our data, we sampled structures directly from known biological assemblies. Currently, this top-down strategy is exclusively applied to protein structures, with plans to extend coverage to DNA and RNA assemblies in future iterations. Using the AlphaFold Protein Structure Database (AFDB)~\cite{afdb} as a source, we randomly selected protein structures and solvated them in explicit water. We then extracted spherical clusters centered on specific residues, capturing the local environment including neighboring residues and water molecules. The truncated boundaries were chemically capped (e.g., avoiding bare H$^+$ or O$^-$) to maintain physical realism. This approach, similar to the methodology used in GEMS~\cite{gems}, ensures that our model learns from geometries highly relevant to real-world biological systems, capturing the complex, non-covalent interaction networks found in native protein folds. The overall bottom-up/top-down strategy is summarized in Figure~\ref{fig:framework}.

\subsubsection{Computational Methodology}
The OMol25 dataset utilizes the $\omega$B97M-V functional~\cite{wb97mv} with the def2-TZVPD basis set~\cite{weigend2005balanced}, a combination that delivers exceptional accuracy for small molecules. However, applying this level of theory to large biological systems directly exacerbates the ``scale-accuracy gap'' due to two major challenges:
\begin{enumerate}
  \item \textbf{Computational Cost:} DFT scaling ranges from $\mathcal{O}(N^3)$ to $\mathcal{O}(N^4)$. As an illustrative example, scaling a 50-atom system to 450 atoms (a 9$\times$ increase) would result in a prohibitive $>700\times$ increase in computational cost.
  \item \textbf{Convergence Issues:} The diffuse functions in the def2-TZVPD basis set often lead to self-consistent field (SCF) convergence failures in large, dense systems. We observed that for systems around 600 atoms, the convergence rate dropped from $>90\%$ (with def2-TZVP) to $<20\%$ (with def2-TZVPD).
\end{enumerate}

To address these issues while maintaining high fidelity, we adopted a pragmatic approach:
\begin{itemize}
  \item \textbf{Functional Selection:} We utilized $\omega$B97M-D3~\cite{wb97md3}, replacing the expensive VV10 non-local correlation with the D3 dispersion correction. This offers comparable accuracy for biological interactions at a significantly reduced cost.
  \item \textbf{Mixed Basis Set Strategy:} We employed a mixed basis set scheme. For hydrogen atoms and metal ions, we used the def2-TZVP basis set, while other elements retained the def2-TZVPD basis. This reduces the number of basis functions by approximately one-third and improves SCF convergence rates from $<20\%$ to $>60\%$ for large systems.
  \item \textbf{Multi-Fidelity Data:} To maximize data coverage, we generated a portion of the dataset using the def2-SVP basis set. Although smaller, def2-SVP is 50--100$\times$ faster than def2-TZVPD on large systems and sufficient for capturing many non-reactive biological interactions. This allowed us to expand the dataset size by $10\times$ with less than 10\% of the total compute budget (internal accounting).
\end{itemize}

\subsubsection{Data Generation Pipeline}
\begin{figure}[H]
  \centering
  \includegraphics[width=1.0\linewidth]{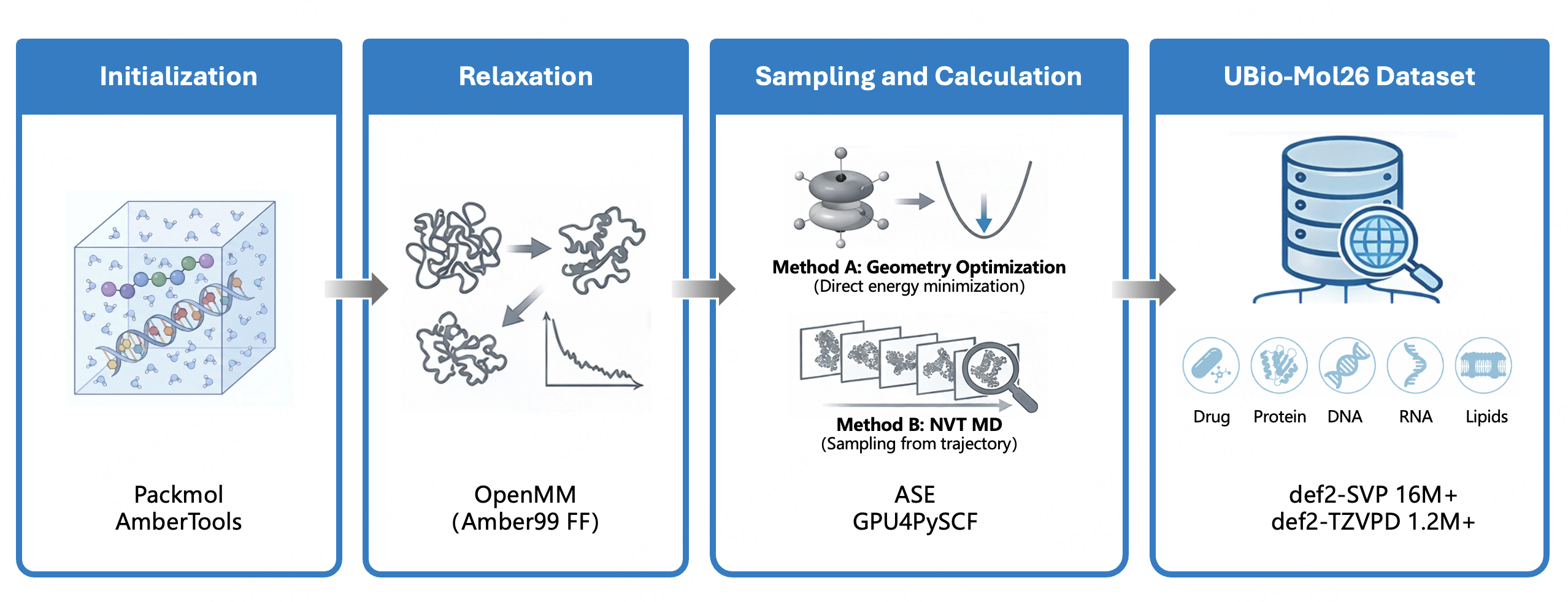}
  \caption{\textbf{Data Generation Pipeline.} (1) \textit{Assembly:} Solutes and solvents are packed using Packmol and parameterized with AmberTools. (2) \textit{Relaxation:} Steric clashes are removed via OpenMM minimization. (3) \textit{Calculation:} High-fidelity DFT labels are generated using GPU4PySCF, employing both direct optimization and MD sampling strategies to maximize structural diversity.}
  \label{fig:data_pipeline}
\end{figure}

The data generation workflow, depicted in Figure \ref{fig:data_pipeline}, consists of two main stages:

\paragraph{1. Initial Configuration Preparation.}
The pipeline begins with the assembly of solvated systems. We used \texttt{Packmol}~\cite{packmol} to pack solute molecules (e.g., proteins, drugs) and solvent molecules into a simulation box with defined density. \texttt{AmberTools} (\texttt{tleap})~\cite{ambertools} was then employed to generate topology files and assign initial force field parameters (Amber99). To resolve high-energy steric clashes introduced during packing, we performed energy minimization and short equilibration using \texttt{OpenMM}~\cite{openmm}. Finally, \texttt{MDAnalysis}~\cite{mdanalysis} was used to carve out spherical clusters or select specific regions of interest, ensuring charge neutrality by adding counterions where necessary.

\paragraph{2. High-Fidelity Calculation.}
The prepared configurations were processed using \texttt{GPU4PySCF}~\cite{gpu4pyscf}, a GPU-accelerated quantum chemistry package. We implemented a dual-branch strategy driven by the \texttt{ASE} (Atomic Simulation Environment)~\cite{ase} interface:
\begin{itemize}
  \item \textbf{DFT Optimization:} A subset of configurations underwent direct geometry optimization at the $\omega$B97M-D3/def2-SVP (or def2-TZVPD) level. This step provides high-quality equilibrium structures and diverse off-equilibrium geometries along the optimization trajectory.
  \item \textbf{MD Sampling:} To explore phase space more broadly, we performed short ab initio molecular dynamics (AIMD) or ML-driven MD using a pilot model. Snapshots were sampled from these trajectories and re-computed with high-precision DFT to serve as labeled training data.
\end{itemize}
The final dataset comprises a 4:1 ratio of optimization trajectories to MD-sampled structures. All DFT calculations used a tight convergence threshold of $10^{-6}$ Hartree via the ASE interface.

\subsubsection{Data Analysis and Visualization}
With the 17 million configurations generated through the pipeline above, we performed a comprehensive analysis to ensure the quality and diversity of UBio-Mol26, focusing on its composition, structural variety, and physical distributions.

\paragraph{Compositional Analysis.}
The 17 million configurations in UBio-Mol26 are distributed across several key biological categories to provide a balanced representation of the life sciences domain. The dataset is primarily composed of two subsets based on the level of theory: the def2-SVP subset, containing approximately 16.1 million configurations with an average system size of 447 atoms, and the high-fidelity def2-TZVPD subset, comprising about 1.27 million configurations with an average size of 248 atoms. The distribution of system sizes across the entire dataset is shown in Figure~\ref{fig:data_stats}a.

As shown in Figure~\ref{fig:data_stats}b, the dataset is further categorized by biological domain:

\begin{figure}[H]
  \centering
  \begin{subfigure}[b]{0.48\linewidth}
    \includegraphics[width=\linewidth]{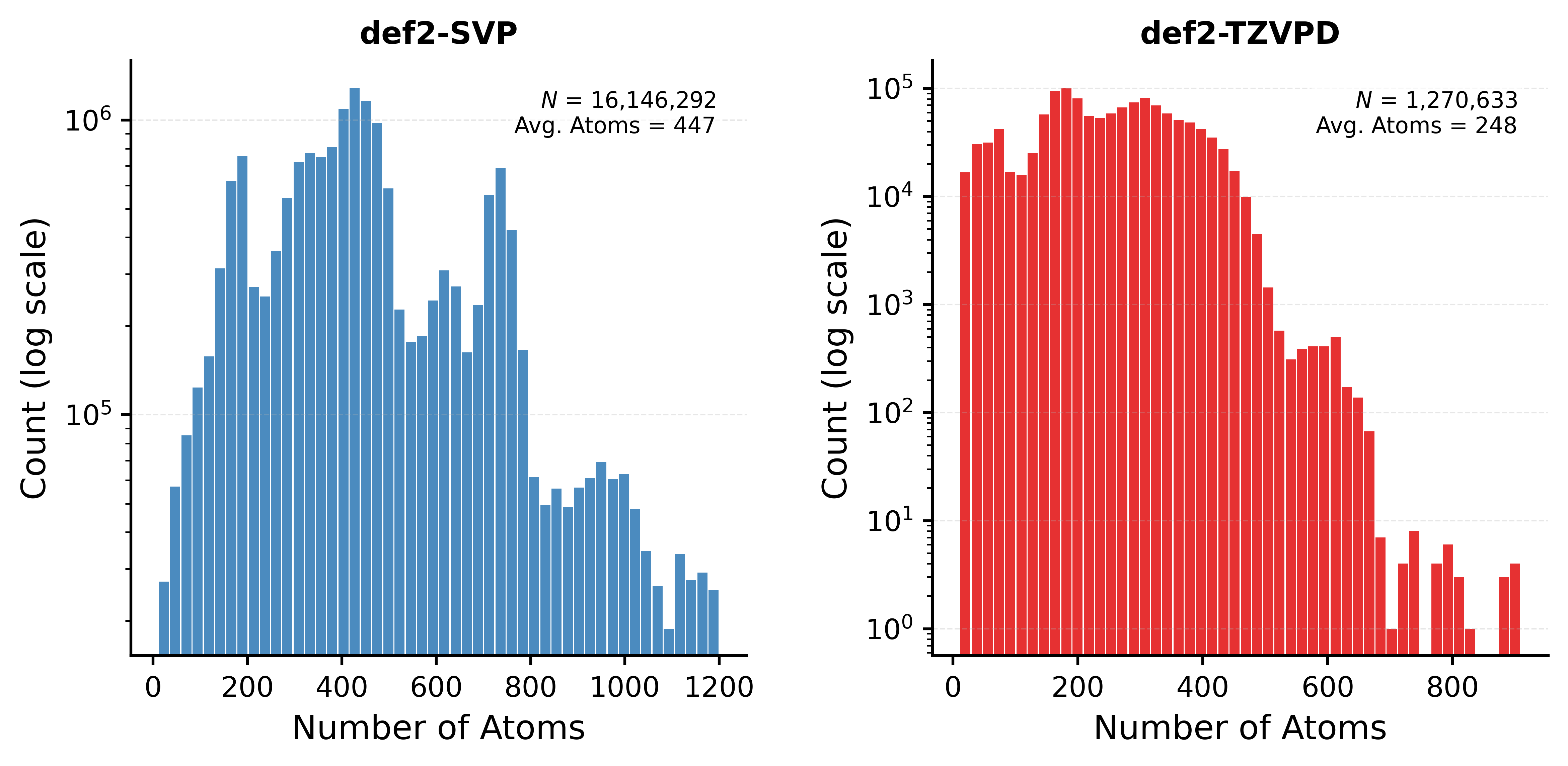}
    \caption{Distribution of system sizes}
    \label{fig:atom_counts}
  \end{subfigure}
  \hfill
  \begin{subfigure}[b]{0.48\linewidth}
    \includegraphics[width=\linewidth]{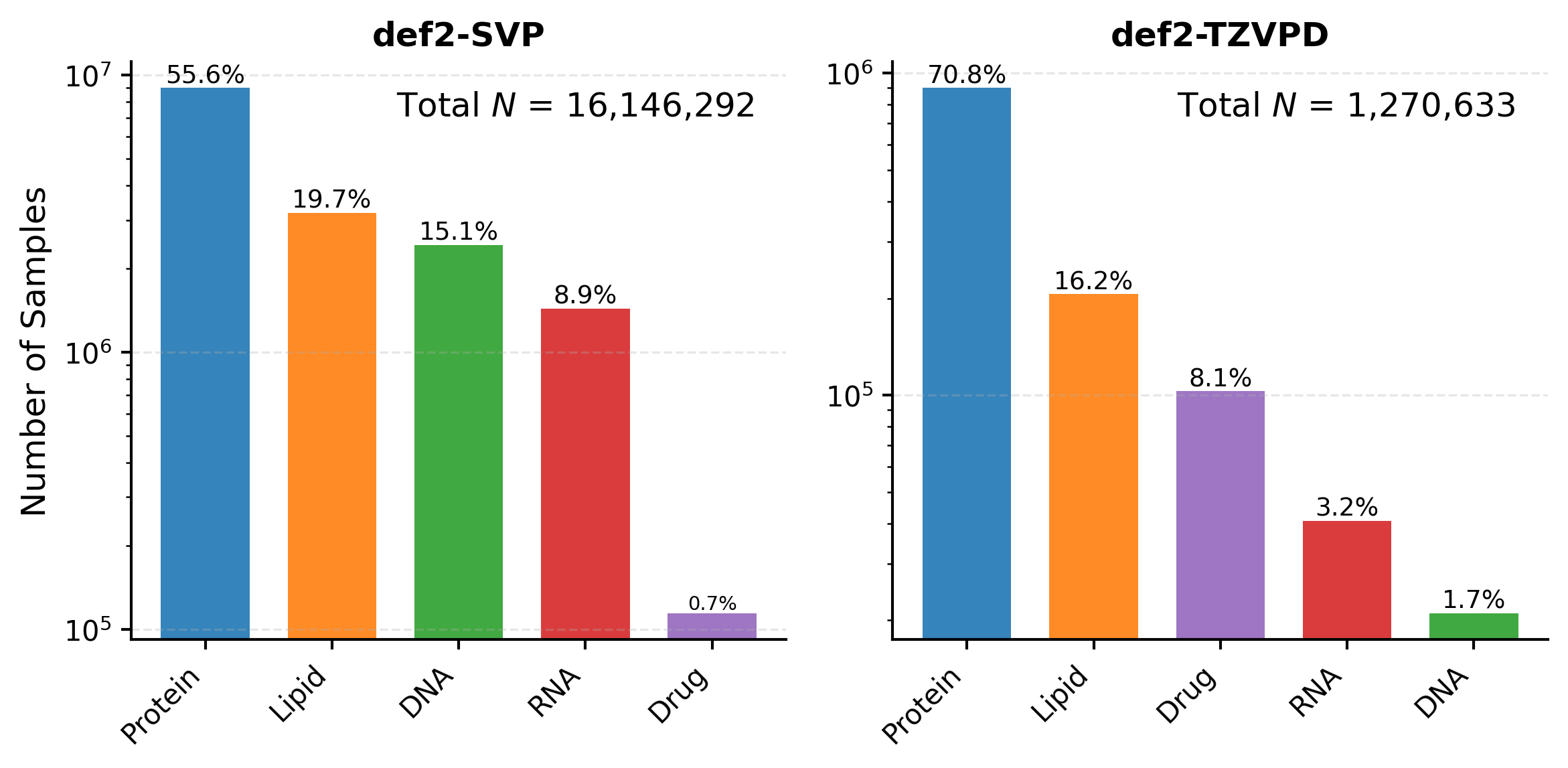}
    \caption{Dataset Composition}
    \label{fig:proportion_all}
  \end{subfigure}
  \caption{\textbf{Statistical overview of UBio-Mol26.} (a) Histogram of atom counts highlighting the extensive coverage of large-scale biological systems. (b) Distribution of configurations across different biological categories.}
  \label{fig:data_stats}
\end{figure}
\begin{itemize}
  \item \textbf{Proteins:} This category forms the backbone of our dataset, constructed through a dual approach. It includes exhaustively enumerated solvated tripeptides to capture fundamental backbone dynamics (bottom-up), as well as diverse residue-centered clusters extracted from the AlphaFold Database (top-down). These clusters are immersed in explicit water spheres to faithfully represent the native aqueous environment of biological systems.
  \item \textbf{Drug-like Molecules:} Comprising small molecules sourced from diverse chemical libraries. These are simulated as solvated clusters, often including nearby water molecules to capture critical solute-solvent interactions essential for binding affinity and pharmacokinetics.
  \item \textbf{DNA/RNA:} To support genomic modeling, we include solvated clusters of nucleic acid fragments. These range from fundamental base-pair assemblies to short strand segments, ensuring the model captures the specific electrostatics and hydrogen bonding patterns of genetic material in solution.
  \item \textbf{Lipid Bilayers:} Representing the cellular boundary, this subset consists of solvated membrane patches. These clusters capture the unique hydrophobic-hydrophilic interfaces and lipid tail dynamics characteristic of biological membranes.
\end{itemize}

\paragraph{Structural and Chemical Diversity.}
To demonstrate the unique biological coverage of UBio-Mol26, we conducted a systematic comparison with OMol25 at both the global structural level and the local chemical level.

\begin{figure}[H]
  \centering
  \includegraphics[width=1.0\linewidth]{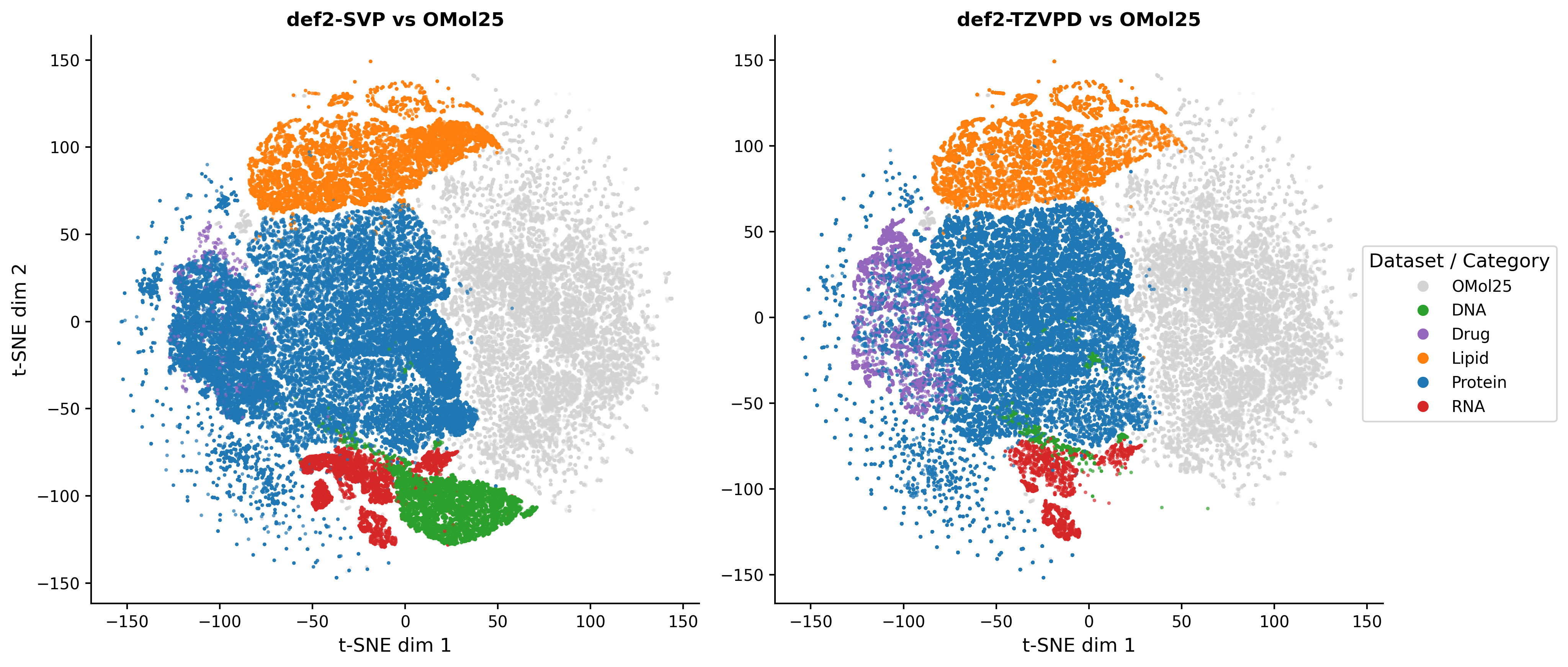}
  \caption{\textbf{t-SNE comparison of UBio-Mol26 and OMol25.} Each panel shows the distribution of 1M sampled configurations in the t-SNE-reduced feature space, categorized by biological domain and basis set (def2-SVP vs def2-TZVPD). The gray background represents OMol25, highlighting the complementary coverage of UBio-Mol26 in the macromolecular regime.}
  \label{fig:tsne_comp}
\end{figure}

\textbf{Global Coverage via t-SNE.} We first visualized the structural distribution of the datasets using t-SNE dimensionality reduction on elemental composition features. We sampled 1 million configurations from each subset (OMol25, def2-SVP, and def2-TZVPD). As shown in Figure~\ref{fig:tsne_comp}, the projection reveals a clear separation: OMol25 (gray background) is concentrated in the small-molecule regime, while UBio-Mol26 extends into broad, previously unexplored regions corresponding to complex macromolecules. The distinct clusters for proteins, DNA/RNA, and lipids confirm that our dataset physically covers the biological domains missing from general-purpose benchmarks.

\begin{figure}[H]
  \centering
  \includegraphics[width=1.0\linewidth]{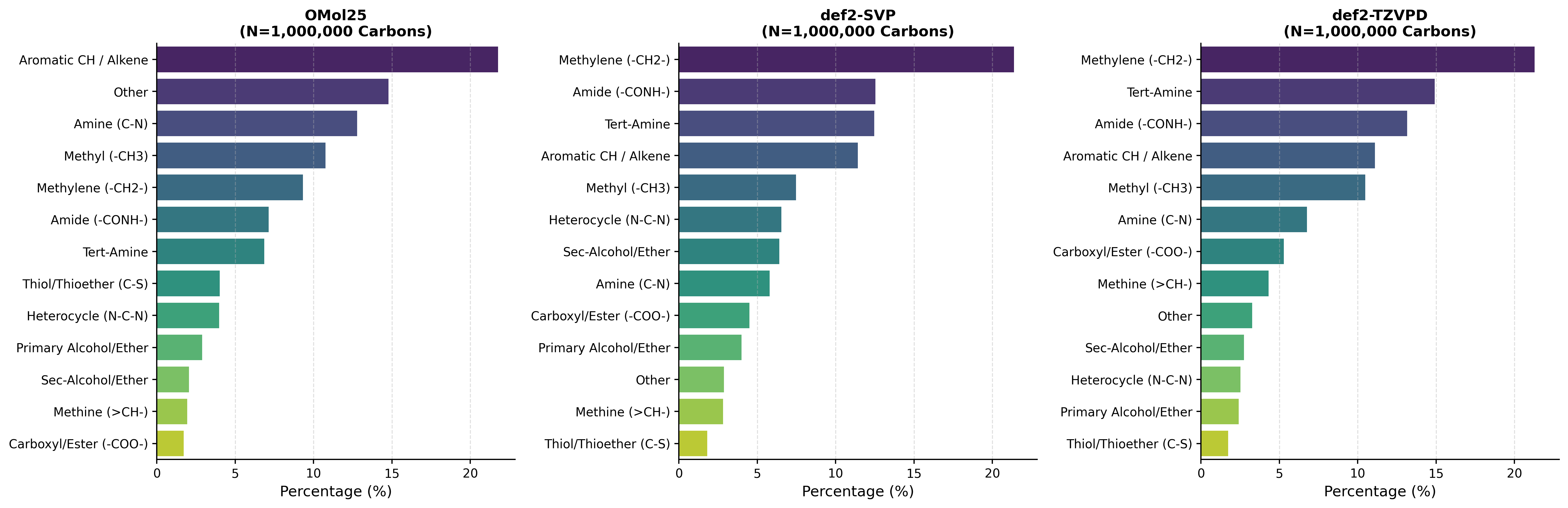}
  \caption{\textbf{Distribution of Carbon chemical environments.} Comparison of functional group frequencies for one million Carbon atoms sampled from OMol25 (left) and UBio-Mol26 subsets (def2-SVP, middle; def2-TZVPD, right). UBio-Mol26 exhibits a significantly higher proportion of methylene and amide groups, consistent with its emphasis on proteins and lipids, whereas OMol25 is enriched in aromatic groups typical of small molecules.}
  \label{fig:organic_dist}
\end{figure}

\textbf{Local Chemical Environments.} To explain the chemical basis of this structural expansion, we analyzed the local functional groups of Carbon atoms (Figure~\ref{fig:organic_dist}). Carbon serves as the scaffold for biological complexity. Our analysis reveals a fundamental compositional shift: OMol25, designed as a dataset for training general-purpose foundation models, is dominated by Aromatic CH / Alkene groups ($\sim$22\%), typical of synthetic drug-like small molecules. In contrast, UBio-Mol26 is specifically tailored for biological systems, evidenced by its enrichment in Methylene (-CH$_2$-) ($\sim$22\%) and Amide (-CONH-) groups—the defining structural motifs of protein backbones and lipid membranes. This data quantitatively confirms that UBio-Mol26 does not simply add ``more atoms,'' but captures the specific chemical environments essential for modeling the intricate physics of biological life.

\begin{figure}[H]
  \centering
  \begin{subfigure}[b]{0.75\linewidth}
    \includegraphics[width=\linewidth]{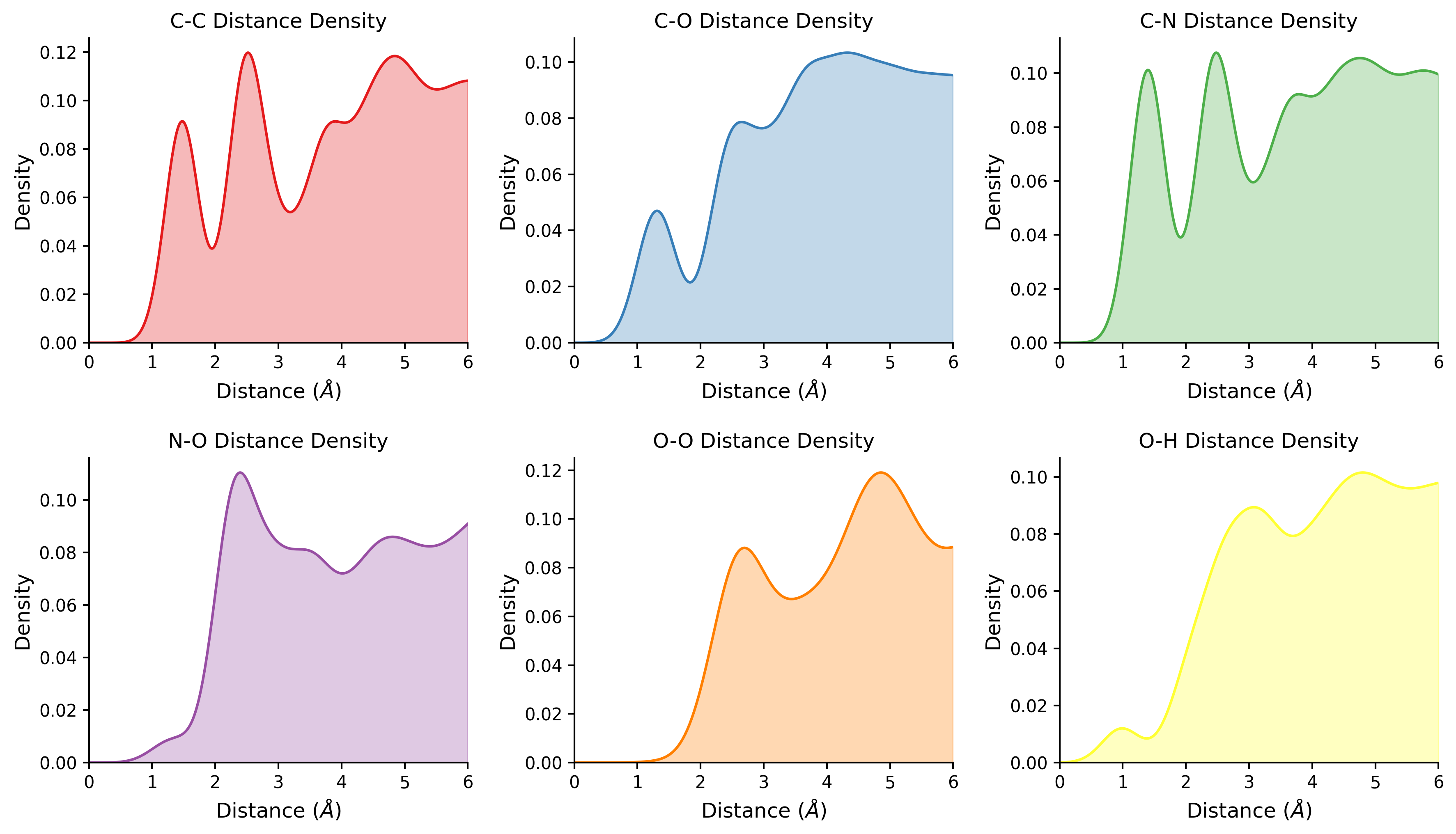}
    \caption{OMol25}
    \label{fig:pair_dist_omol}
  \end{subfigure}
  \vspace{0.5cm}
  \begin{subfigure}[b]{0.75\linewidth}
    \includegraphics[width=\linewidth]{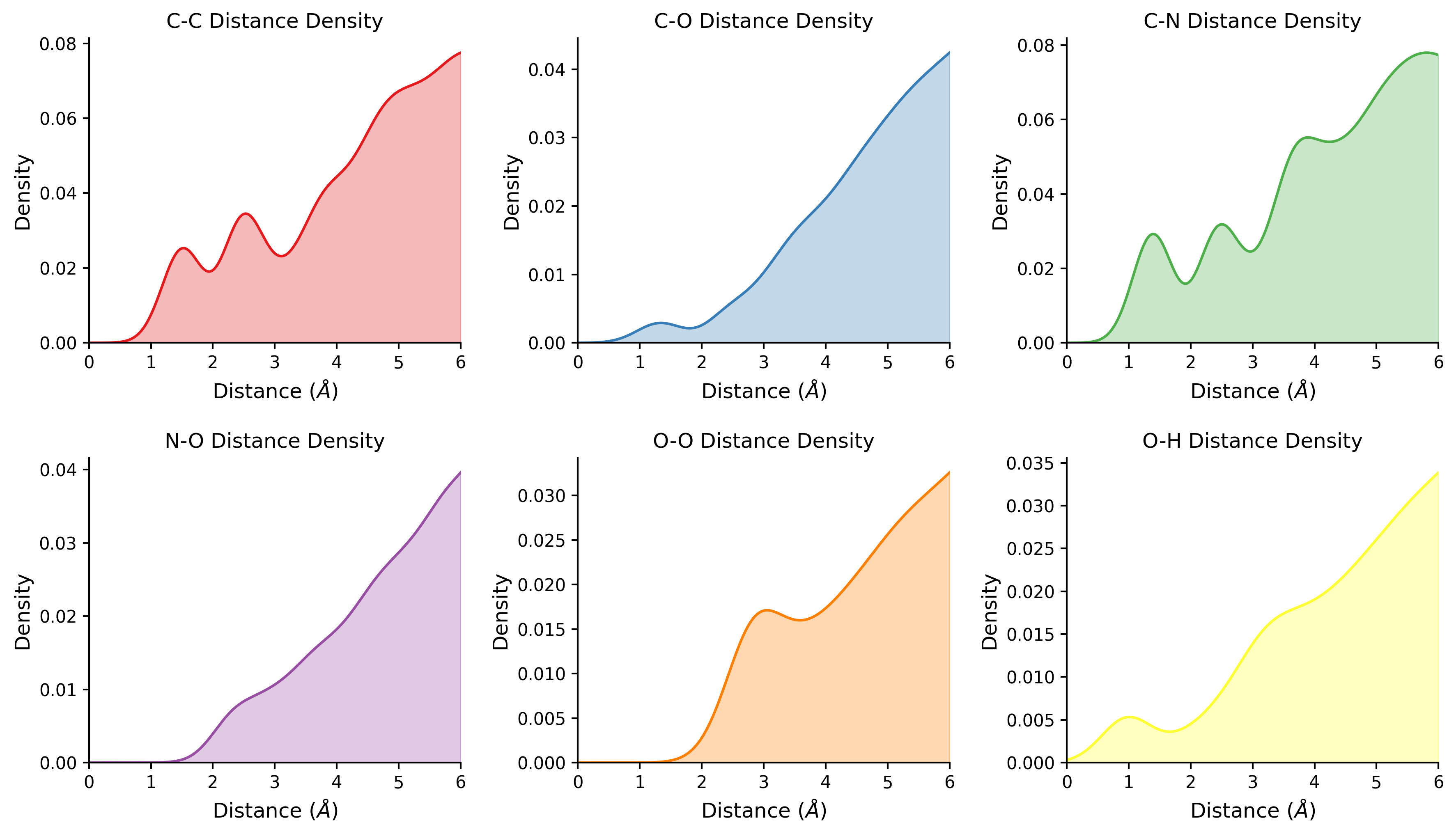}
    \caption{UBio-Mol26 (def2-SVP)}
    \label{fig:pair_dist_svp}
  \end{subfigure}
  \vspace{0.5cm}
  \begin{subfigure}[b]{0.75\linewidth}
    \includegraphics[width=\linewidth]{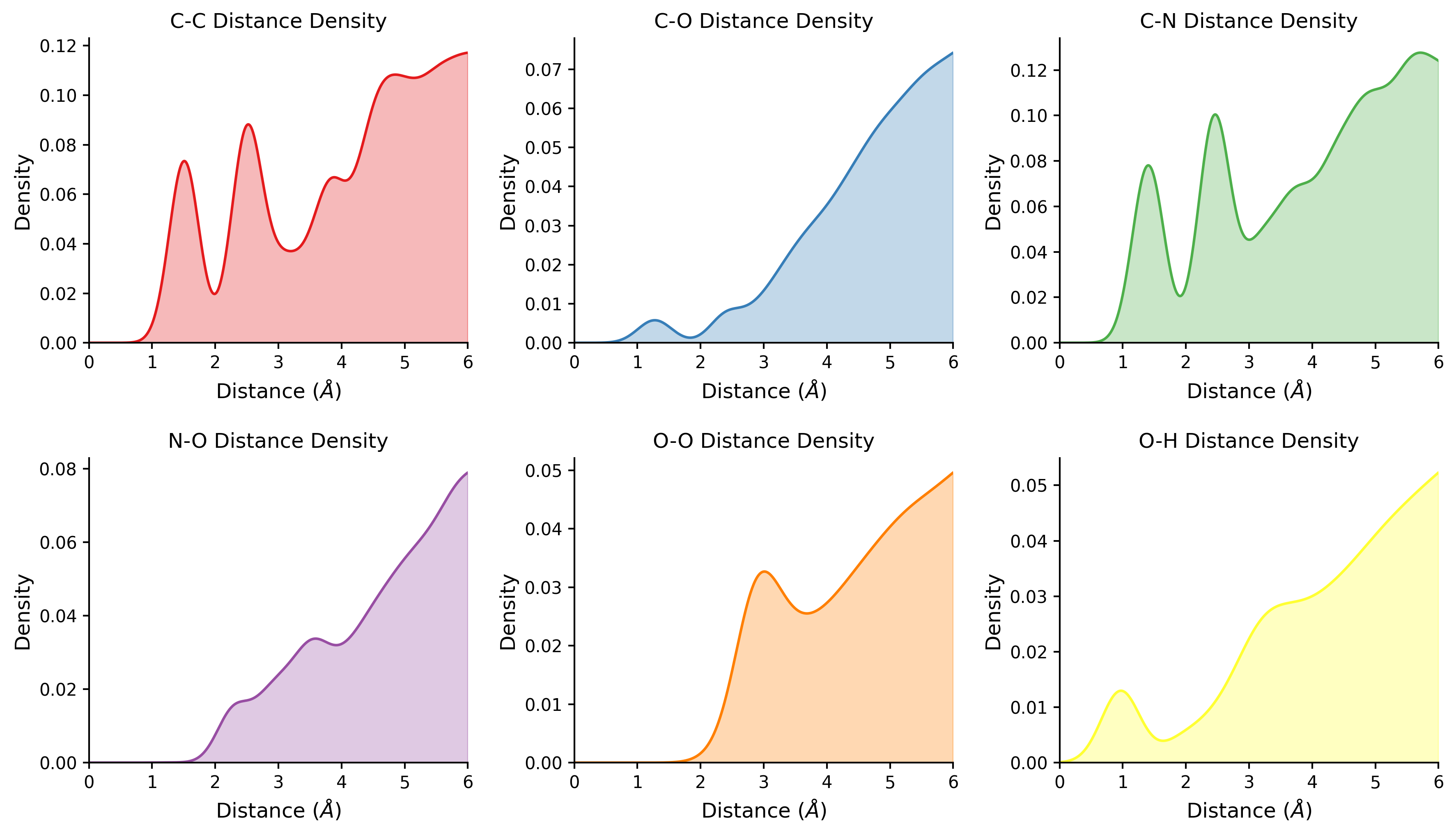}
    \caption{UBio-Mol26 (def2-TZVPD)}
    \label{fig:pair_dist_tzvpd}
  \end{subfigure}
  \caption{\textbf{Pair distance distributions for C, N, O, H.} Comparison of interatomic distance distributions across OMol25 and UBio-Mol26 subsets. UBio-Mol26 captures significantly longer-range structural information essential for biological macromolecules.}
  \label{fig:pair_distances}
\end{figure}

\paragraph{Physical Distributions.}
The physical fidelity of the dataset is validated through the distribution of interatomic distances. We analyzed the pair distance distributions for common biological elements (C, N, O, H) across OMol25, the def2-SVP subset, and the def2-TZVPD subset. As shown in Figure~\ref{fig:pair_distances}, all three datasets exhibit consistent distributions in the short-range regime (3--4~\AA), with peaks aligning closely with empirical chemical bond lengths. This indicates that all subsets accurately capture fundamental local bonding geometries. However, UBio-Mol26 (def2-SVP and def2-TZVPD) shows a significantly broader distribution compared to OMol25, reflecting the larger system sizes and more complex macromolecular environments. In contrast, the distribution for OMol25 begins to decline sharply beyond 5--6~\AA, highlighting its limitation in representing long-range structural correlations.

\begin{figure}[H]
  \centering
  \begin{subfigure}[b]{0.95\linewidth}
    \includegraphics[width=\linewidth]{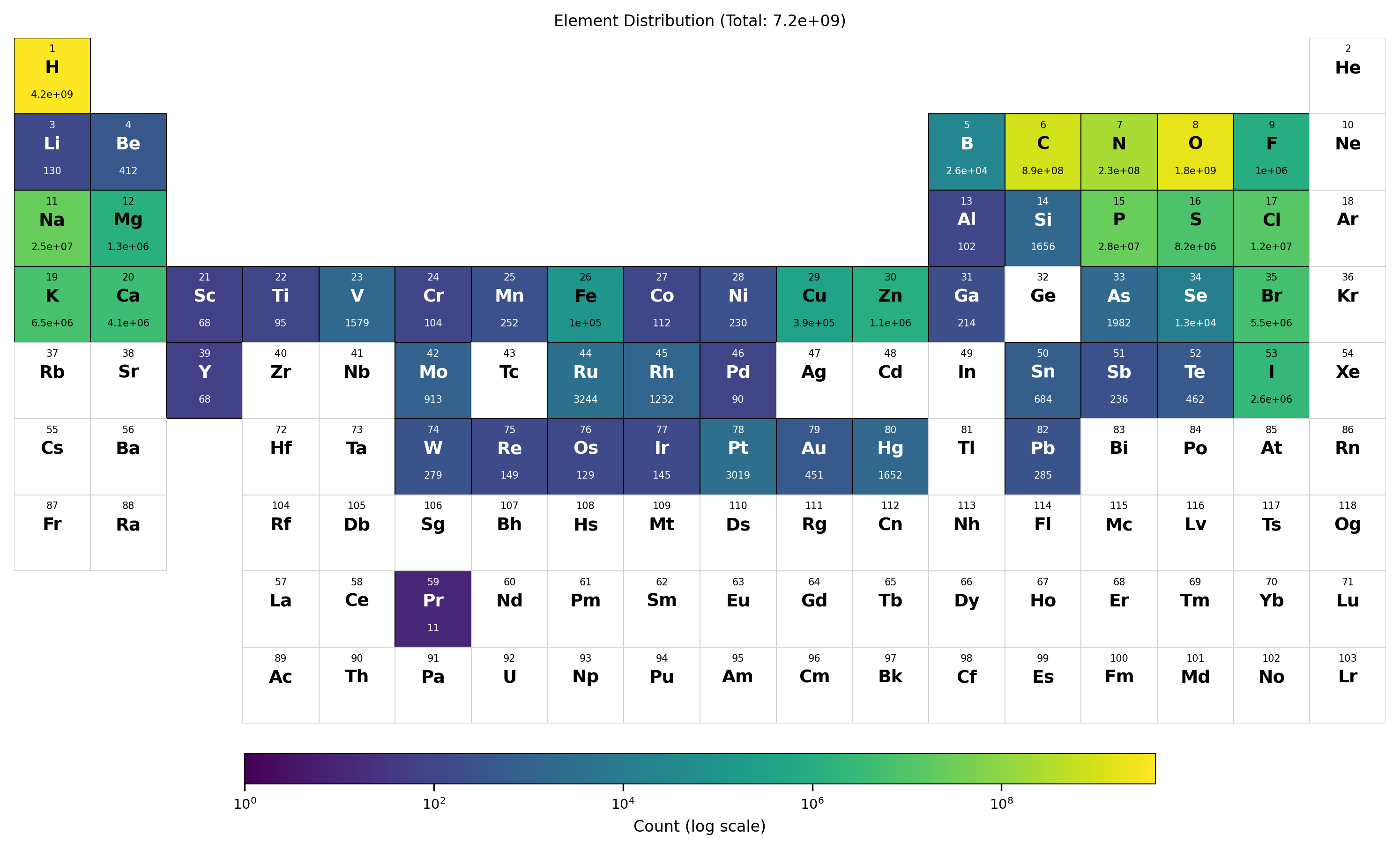}
    \caption{def2-SVP Subset}
    \label{fig:atom_dist_svp}
  \end{subfigure}
  \vspace{0.5cm}
  \begin{subfigure}[b]{0.95\linewidth}
    \includegraphics[width=\linewidth]{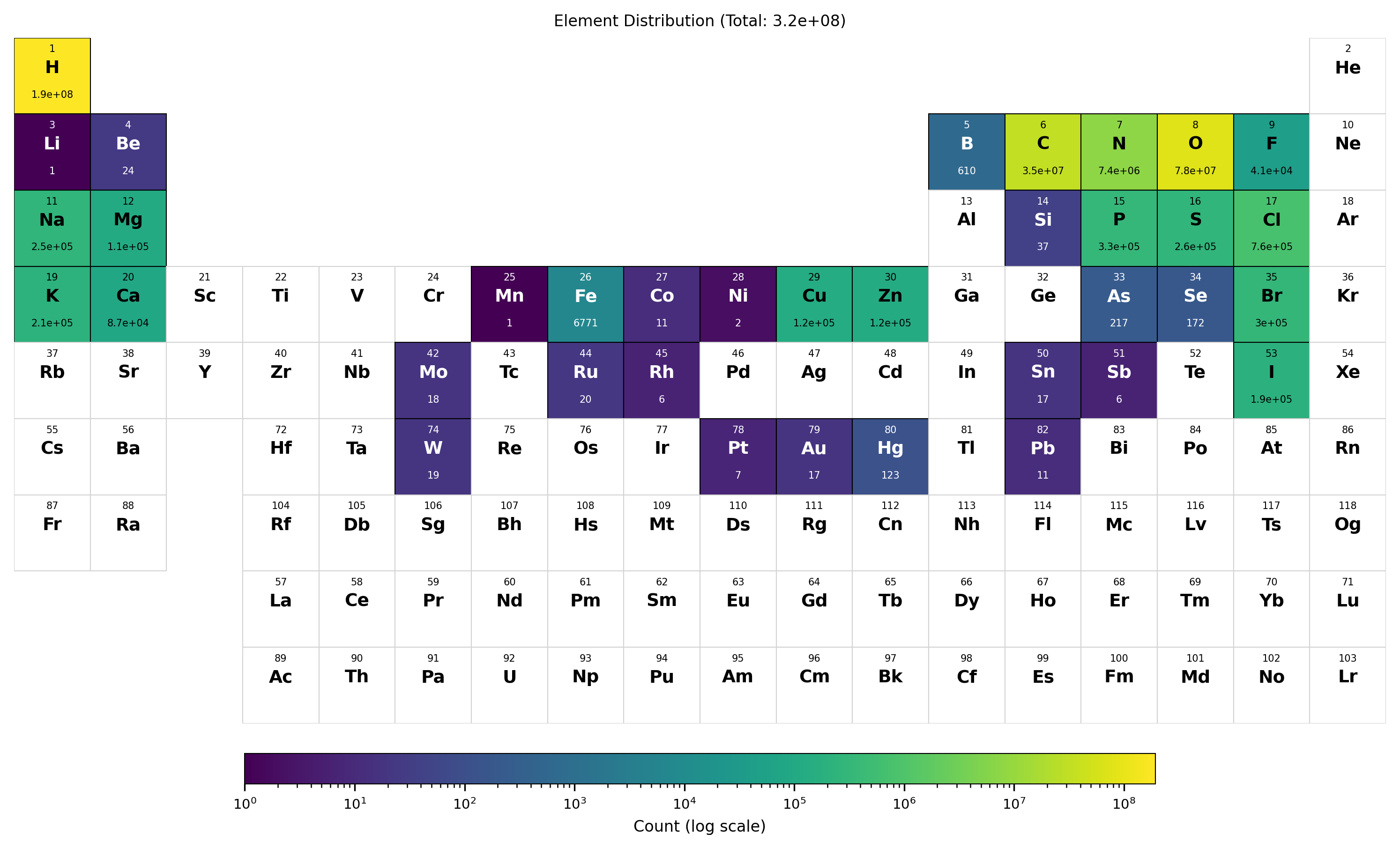}
    \caption{def2-TZVPD Subset}
    \label{fig:atom_dist_tzvpd}
  \end{subfigure}
  \caption{\textbf{Elemental distribution across the periodic table.} Heatmaps showing the frequency of elements included in the def2-SVP and def2-TZVPD subsets of UBio-Mol26, demonstrating comprehensive coverage of biologically relevant elements and trace ions.}
  \label{fig:atom_dist}
\end{figure}

\paragraph{Elemental Coverage.}
UBio-Mol26 provides extensive coverage of the periodic table. We visualized the occurrence frequency of elements for both the def2-SVP and def2-TZVPD subsets (Figure~\ref{fig:atom_dist}). The dataset comprehensively covers the core elements of life (C, H, N, O, P, S) as well as a wide range of trace ions and metals (e.g., Mg$^{2+}$, Zn$^{2+}$, Fe$^{2+/3+}$) that are critical for biological function and structural stability. This broad elemental diversity ensures the model's generalizability across diverse biochemical environments.

\subsection{Model Architecture}
\label{sec:model}
We first summarize the representation, architectural, and computational principles underlying our approach.
Our objective is to construct an $\mathrm{SO}(3)$-equivariant molecular model that is both scalable to large, heterogeneous systems and efficiently realizable on modern hardware.
To this end, we build upon the node-centric factorization of E2Former and extend it with a long--short range (LSR) architecture and a hardware-aware execution model.
And then, we note that although the resulting message passing scheme is linear-scaling in principle, its practical performance is limited by dense tensor contractions and edge-wise memory materialization.
This gap motivates the hardware-efficient extensions introduced in \S\ref{subsec:v2}.

\subsubsection{Preliminaries and Method Overview}
\label{sec:prelim_method_overview}

\paragraph{Notion.}
Throughout the paper, $\ell \ge 0$ denotes the angular momentum degree and
$m \in \{-\ell,\ldots,\ell\}$ the associated magnetic index.
All spherical harmonics are real-valued.
Vectors in $\mathbb{R}^3$ are denoted by $\vec r$.
Irreducible components of degree $\ell$ are indicated by the superscript $(\ell)$,
and the operator $[\cdot]^{(\ell)}$ extracts the $\ell$-th order component from a
representation or tensor product.
Clebsch--Gordan tensor products are denoted by $\otimes$, while $\otimes^{6j}$
refers to Wigner-$6j$ recoupling, which will be introduced later.

\paragraph{Graph and $\mathrm{SO}(3)$-equivariant node features.}
We represent a molecular system as a graph
$\mathcal{G}=(\mathcal{V},\mathcal{E})$ with $N = |\mathcal{V}|$ nodes.
Each node $i \in \mathcal{V}$ is associated with a position
$\vec r_i \in \mathbb{R}^3$ and an equivariant feature vector $\mathbf h_i$.
Relative displacements are defined as $\vec r_{ij} = \vec r_j - \vec r_i$.

Node features are modeled as a direct sum of irreducible representations of
$\mathrm{SO}(3)$:
\begin{equation}
  \mathbf h_i = \bigoplus_{\ell=0}^{L} \mathbf h_i^{(\ell)}, \qquad
  \mathbf h_i^{(\ell)} \in \mathbb{R}^{C_\ell \times (2\ell+1)},
\end{equation}
where $C_\ell$ denotes the number of channels at degree $\ell$.
Under a rotation $R \in \mathrm{SO}(3)$, each component transforms equivariantly as
\begin{equation}
  \mathbf h_i^{(\ell)} \mapsto D^{(\ell)}(R)\,\mathbf h_i^{(\ell)},
\end{equation}
with $D^{(\ell)}(R)$ the Wigner-$D$ matrix.
In practice, scalar features ($\ell=0$) are initialized from node-type embeddings,
while higher-order components are initialized to zero, allowing equivariant
structure to emerge through message passing.

\paragraph{Geometric encoding via solid spherical harmonics.}
Geometric relationships between atoms are encoded using real solid spherical harmonics
$\mathcal{R}^{(\ell)}(\vec r) \in \mathbb{R}^{2\ell+1}$, defined as
\begin{equation}
  \mathcal{R}^{(\ell)}_m(\vec r)
  =
  \|\vec r\|^\ell\,
  Y_{\ell,m}\!\left(\frac{\vec r}{\|\vec r\|}\right),
\end{equation}
where $Y_{\ell,m}$ denote real spherical harmonics and $\|\vec r\|$ is the Euclidean
norm.
This representation jointly captures angular information and transforms
as an $\ell$-th order irrep under $\mathrm{SO}(3)$ rotations, making it a natural
building block for equivariant interactions.

\paragraph{Clebsch--Gordan tensor products.}
To construct interactions between features of different degrees while preserving
equivariance, we employ Clebsch--Gordan (CG) tensor products.
Given irreps $\mathbf u^{(\ell_1)}$ and $\mathbf v^{(\ell_2)}$, their coupling into
an output irrep of degree $\ell_{\mathrm{out}}$ is defined as
\begin{equation}
  \label{eq:cg_product_report}
  \left(
    \mathbf u^{(\ell_1)} \otimes \mathbf v^{(\ell_2)}
  \right)^{(\ell_{\mathrm{out}})}_m
  =
  \sum_{m_1,m_2}
  C^{\ell_{\mathrm{out}},m}_{\ell_1,m_1;\ell_2,m_2}\,
  u^{(\ell_1)}_{m_1}\, v^{(\ell_2)}_{m_2},
\end{equation}
where $C^{\cdot}_{\cdot}$ are Clebsch--Gordan coefficients.
This operation guarantees that the resulting features transform equivariantly
under rotations.
For clarity, parity labels are omitted throughout this work.

\paragraph{Equivariant attention and message aggregation.}
Equivariant message passing is implemented through an attention mechanism that
aggregates information from neighboring nodes while maintaining rotational
equivariance.
For node $i$, the aggregated message is given by
\begin{equation}
  \label{eq:generic_attn}
  \mathbf m_i
  =
  \sum_{j \in \mathcal{N}(i)} \alpha_{ij}\,\mathbf m_{ij},
  \qquad
  \mathbf m_{ij}
  :=
  \mathbf h_j \otimes \frac{\mathcal{R}^{(\ell)}(\vec r_{ij})}{\|\vec r_{ij}\|} ,
\end{equation}
where $\alpha_{ij}$ is a scalar, rotation-invariant attention weight and
$\mathcal{N}(i)$ denotes the neighbor set of node $i$.
This formulation serves as a generic template for equivariant attention, upon
which we build our node-centric and hardware-efficient implementation.

\paragraph{Long--short range modeling.}
Molecular interactions span multiple spatial scales.
Strong many-body interactions are predominantly local, while long-range effects
such as electrostatics, polarization, and through-space correlation decay slowly
with distance.
To reflect this physical structure, we adopt a long--short range (LSR) architecture.
The short-range module captures local interactions within a radius graph
($r_{\text{short}} \approx 5\,\text{\AA}$) using a stack of E2Former layers with
Wigner-$6j$--based equivariant attention.
This module learns high-resolution local potential energy surfaces with linear
scaling in the number of atoms.
Long-range interactions are modeled through a bipartite atom--fragment graph,
in which each atom attends to a small set of fragment nodes representing distant
regions of the system.
This design extends the receptive field to $r_{\text{long}} \approx 15\,\text{\AA}$
without forming a fully connected atomic graph, enabling efficient propagation of
non-local information while preserving $\mathrm{SO}(3)$ equivariance:
\begin{equation}
  \mathbf m^{\text{S}}_i
  =
  \sum_{j \in \mathcal{N}_{r_{\text{short}}}(i)} \alpha_{ij}\,\mathbf m_{ij},
  \qquad
  \mathbf m^{\text{L}}_i
  =
  \sum_{j \in \mathcal{N}_{r_{\text{long}}}(i)} \alpha_{ij}\,\mathbf m_{ij}.
\end{equation}

\paragraph{Late fusion and property prediction.}
We combine multi-scale information through \emph{late fusion} of invariant features.
Specifically, let $\mathbf m^{\text{S},(\ell=0)}_i$ and $\mathbf m^{\text{L},(\ell=0)}_i$
denote the scalar ($\ell=0$) components of the final short- and long-range messages,
respectively.
Restricting fusion to invariant channels ensures rotational invariance of the
energy prediction while allowing equivariant higher-order features to influence
forces through their dependence on atomic positions.
The fused atomic representation is formed by concatenation followed by a
shared multilayer perceptron,
and the total molecular energy is predicted as a sum of atomic contributions. Atomic forces are obtained by analytical differentiation of the predicted energy
with respect to atomic positions,
\begin{equation}
  \widehat{E} = \sum_i\mathrm{MLP}_{\text{fuse}}\!\left(
    [\mathbf m^{\text{S},l=0}_{i} \,\Vert\, \mathbf m^{\text{L},l=0}_{i}]
  \right),
  \qquad
  \widehat{\mathbf F}_i
  =
  -\,\frac{\partial \widehat{E}}{\partial \vec r_i}.
\end{equation}
which guarantees energy--force consistency by construction.

\subsubsection{E2Former-V2: Hardware-Efficient Scaling}
\label{subsec:v2}

\paragraph{E2Former node-centric factorization .}
Applying BLE and Wigner--$6j$ recoupling to the edge message
$\mathbf m_{ij}=\mathbf h_j\otimes \mathcal R^{(\ell)}(\vec r_{ij})$
yields the factorized form:
\begin{equation}
  \label{eq:e2former_final_report}
  \begin{split}
    \mathbf m_i^{(\ell_{\mathrm{out}})}
    =
    \sum_{u=0}^{\ell} (-1)^{\ell-u}\binom{\ell}{u}\,
    \bigg[
      \underbrace{\mathcal R^{(u)}(\vec r_i)}_{\text{target term}}
      \ \otimes_{6j}\
      \bigg(
        \sum_{j\in\mathcal N(i)} \frac{\alpha_{ij}}{\|\vec r_{ij}\|^l}\,
        \underbrace{\big(\mathbf h_j\otimes \mathcal R^{(\ell-u)}(\vec r_j)\big)}_{\text{source term}}
      \bigg)
    \bigg]^{(\ell_{\mathrm{out}})}.
  \end{split}
\end{equation}

Our attention mechanism realizes the node-centric factorization \cite{li2025e2former} by strictly confining geometric information to the node-wise value path. The message passing decomposes into three stages: (i) \textit{source-term preparation}, where value features are pre-coupled with local spherical harmonics at source nodes; (ii) \textit{weighted aggregation}, where pre-computed terms are transmitted using scalar attention weights $\alpha_{ij}$; and (iii) \textit{target-term coupling}, where the aggregated message is coupled with target node spherical harmonics. Critically, the scalar $\alpha_{ij}$ represents the \textit{only} edge-dependent term, enabling memory-efficient computation.


\begin{figure}[H]
  \centering
  \includegraphics[width=0.95\linewidth]{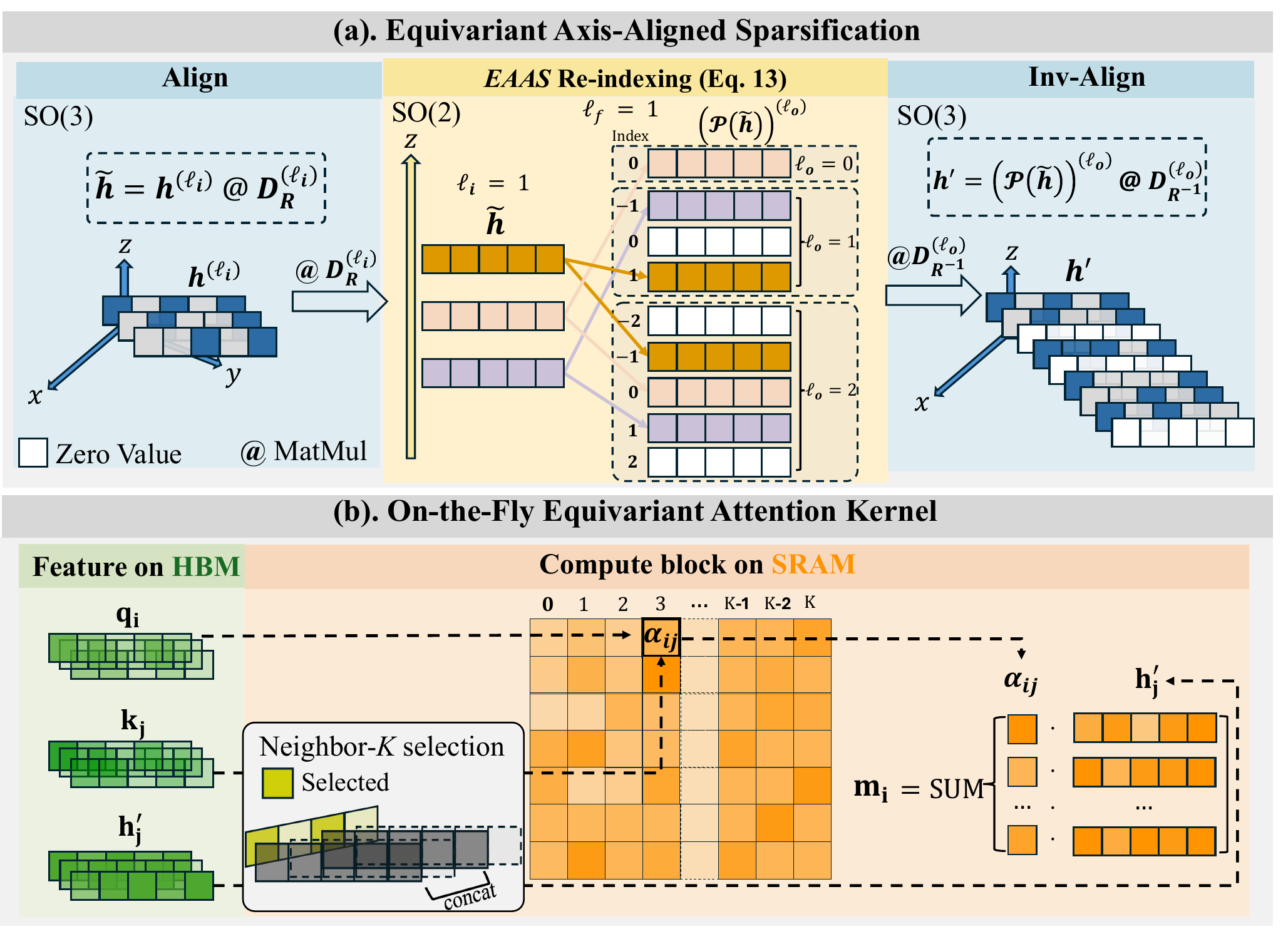}
  \caption{\textbf{E2Former-V2 key components.} We propose Equivariant Axis-Aligned Sparsification (EAAS) to simplify tensor couplings through $\mathrm{SO}(2)$ re-indexing and utilize a fused on-the-fly equivariant kernel for memory-efficient, edge-free attention computation.}
  \label{fig:e2former_v2}
\end{figure}


\paragraph{Equivariant Axis-Aligned Sparsification (EAAS)}
To resolve the arithmetic bottleneck of dense tensor products, we introduce EAAS, an algebraic reduction that converts dense $\mathrm{SO}(3)$ tensor products into sparse operations while preserving exact equivariance.

The key observation is that solid spherical harmonics become maximally sparse when rotated into an axis-aligned frame: only the $m=0$ component survives. By commuting rotations through the tensor product, dense Clebsch--Gordan couplings collapse into a deterministic re-indexing operation followed by lightweight blockwise linear maps.

For a rotation $R$ that aligns the $z$-axis with vector $\vec{r}$, the $\mathrm{SO}(3)$-equivariant tensor product admits the compact form:
\begin{equation}
  \label{eq:eaas_expanded_report}
  \big(h^{(\ell_i)}\otimes \mathcal R^{(\ell_f)}(\vec r)\big)^{(\ell_o)}_{m_o}
  =
  \big(\mathcal P(\tilde h)\big)^{(\ell_o)}\mathbin{@}D_{R^{-1}}^{(\ell_o)}.
\end{equation}
where $D_R$ are Wigner-$D$ matrices and $\mathcal{P}$ is a sparse re-indexing operator that maps input orders $m_i$ to output orders $m_o$ based on parity rules. This decomposition replaces dense contractions with: (i) re-indexing $\mathcal{P}$ (a memory-efficient permutation), (ii) blockwise linear maps $W^{(\ell)}$ (small $7 \times 7$ matrices for $\ell \leq 3$), and (iii) rotation conjugation via Wigner-$D$ matrices. The entire operation is dramatically faster ($\sim\!6\times$ speedup) than dense Clebsch--Gordan products while preserving \emph{exact} $\mathrm{SO}(3)$ equivariance. Full derivation and parity selection rules are detailed in \cite{huang2026e2former}.

\begin{algorithm}[!t]
  \caption{Fused On-the-Fly Equivariant Attention (Forward Pass, $H=1$)}
  \label{alg:fused_streaming_attn}
  \begin{algorithmic}[1]
    \Require
    $q, k \in \mathbb{R}^{N \times d}$, $h' \in \mathbb{R}^{N \times C}$,
    neighbor set $\mathcal{N}\in\mathbb{R}^{N \times K}$,
    bias $b(r)$, radial scaling $\phi(r)$, scale $\tau = 1/\sqrt{d_k}$
    \Ensure
    Aggregated message $m \in \mathbb{R}^{N \times C}$
    \For{each target atom $i \in \{1,\dots,N\}$ \textbf{in parallel}}
    \State $\mu \gets -\infty$
    \State $z \gets 0$
    \State $\mathbf{A} \gets \mathbf{0}$
    \For{$j\in\mathcal{N}(i)$ }
    \If{$j$ is padding} \State \textbf{continue} \EndIf
    \State $s \gets \tau \cdot q_i^{\top} k_j + b(r_{ij})$
    \State $\mu' \gets \max(\mu, s)$
    \State $z \gets z \cdot e^{\mu - \mu'} + e^{s - \mu'}$
    \State $\mathbf{A} \gets \mathbf{A} \cdot e^{\mu - \mu'} + e^{s - \mu'} \cdot \phi(r_{ij})\, h'_j$
    \State $\mu \gets \mu'$
    \EndFor
    \State $m_i \gets \mathbf{A} / z$
    \EndFor
  \end{algorithmic}
\end{algorithm}

\paragraph{On-the-Fly Equivariant Attention}

While EAAS eliminates the algebraic bottleneck, standard implementations still suffer from the memory wall: materializing attention logits $\alpha_{ij}$ for all edges requires $\mathcal{O}(|\mathcal{E}|)$ HBM access, and gathering neighbor keys/values creates intermediate tensors that consume gigabytes of memory.

As shown in Algorithm \ref{alg:fused_streaming_attn}, we implement a custom GPU kernel using Triton that realizes the edge-free design by computing attention scores ephemerally via online softmax. The kernel evaluates aggregation as a streaming reduction: for each atom $i$, it maintains running accumulators in on-chip SRAM and iterates over neighbors, evaluating inner products on-the-fly without materializing dense edge tensors.

The kernel employs: (i) \textit{atom-major tiling} to keep target geometry and queries in SRAM for maximum data reuse, (ii) \textit{fused source/target operations} computed on-the-fly within the load loop, and (iii) \textit{mixed precision} (FP16/BF16 for weights, FP32 for softmax accumulators) to prevent numerical instability.

\paragraph{Memory reduction.}
By avoiding explicit materialization, this design eliminates dominant memory bottlenecks. Standard implementations require $\mathcal{O}(N \cdot K \cdot H \cdot d)$ for gathered keys and $\mathcal{O}(N \cdot K \cdot H \cdot C)$ for gathered values; our kernel streams these on-the-fly, requiring only ephemeral computation. This enables scaling to million-atom systems with $\sim\!20\times$ TFLOPS improvement compared to standard implementations.


\subsection{Infrastructure \& Training}
\label{sec:training_strategy}
In our setting, the training data span both \textbf{large-scale chemical diversity} and \textbf{highly heterogeneous system sizes}.
The OMol25 dataset contains on the order of 100 million molecular configurations with relatively small systems (50--300 atoms), providing broad coverage of chemical space.
In contrast, the TZVPD and SVP datasets comprise approximately 10 million configurations with substantially larger systems (300--1500 atoms), emphasizing scalability with respect to molecular size.
Training a single model to simultaneously capture both dimensions---chemical diversity and system-size scaling---poses a nontrivial optimization challenge.

Moreover, when the number of atoms varies by more than an order of magnitude across samples, naive graph-count--based batching leads to severe load imbalance, underutilizing hardware resources and limiting scalability.
At this scale, data scheduling is no longer a secondary engineering detail, but a first-order factor that directly impacts training throughput, optimization dynamics, and convergence stability.
To address this challenge, we adopt a distributed training pipeline with a balanced, atom-aware dataloader that aligns the computational unit of the model with the scheduling unit of the training system.

In this section, we provide a detailed description of our multi-stage training strategy and the atom-balanced data loading mechanism.

\subsubsection{Three-Stage Curriculum Learning}

To efficiently scale training while ensuring physical consistency under such dataset heterogeneity, we adopt a \textbf{three-stage training protocol} inspired by curriculum learning.
This staged approach significantly reduces computational cost compared to end-to-end training on the full dataset, while progressively enforcing physical constraints and extending system-size generalization.

\paragraph{Stage 1: Fast Energy Initialization.}
In the first stage, we train a base model exclusively on the OMol25 dataset to rapidly establish a broad and expressive representation of chemical space.
Owing to the relatively small system sizes and the large scale of OMol25, this stage prioritizes throughput and representation learning.

To maximize training efficiency, we employ separate prediction heads for energy and force, and disable the computationally expensive automatic differentiation required for force computation.
Forces are predicted directly by an independent force head, allowing fast convergence of the energy landscape with minimal memory overhead. This direct force-head setting is distinct from the force-only supervision used later, which still computes forces as energy gradients.

\paragraph{Stage 2: Energy--Force Consistency.}
In the second stage, we initialize the model with weights from Stage~1 but discard the independent force prediction head.
Training proceeds using only the energy head, with forces derived strictly as the negative gradient of the predicted energy,
\begin{equation}
  \mathbf{F} = -\nabla_{\mathbf{r}} E,
\end{equation}
via automatic differentiation.

By fine-tuning under this conservative formulation, we explicitly enforce energy conservation and obtain a physically consistent potential energy surface (PES), while preserving the chemical coverage learned during Stage~1.

\paragraph{Stage 3: Mixed-Dataset Fine-Tuning.}
In the final stage, we incorporate our in-house SVP and TZVPD datasets to extend the model's applicability to significantly larger molecular systems.
This stage is performed with a reduced learning rate to improve system-size scaling while maintaining the rich chemical space representation acquired from OMol25.

To accommodate different levels of electronic structure theory, we adopt a dual-head architecture:
\begin{itemize}
  \item \textbf{Head A (SVP):} Dedicated to the def2-SVP subset of UBio-Mol26.
  \item \textbf{Head B (High-Fidelity):} Shared between OMol25 and the def2-TZVPD subset of UBio-Mol26.
\end{itemize}

All forces are computed via automatic differentiation.
To mitigate distribution shifts and systematic energy offsets across datasets, we apply the following data strategies:
\begin{itemize}
  \item \textbf{Data Balancing:} We sample OMol25, def2-SVP, and def2-TZVPD data with a ratio of $8{:}1{:}1$, respectively.
  \item \textbf{TZVPD Filtering and Alignment:} Since the UBio-Mol26 def2-TZVPD subset shares an energy head with OMol25 but employs a different density functional ($\omega$B97M-D3 vs.\ $\omega$B97M-V) and a mixed basis set, an inherent energy offset is observed.
    To address this mismatch:
    \begin{enumerate}
      \item We filter the def2-TZVPD subset using the Stage~2 model, discarding approximately $30\%$ of configurations with low force cosine similarity to ensure compatibility with the OMol25 manifold (threshold tuned to retain the high-similarity majority while stabilizing training).
      \item We train on the remaining def2-TZVPD configurations using \textit{force-only supervision}: forces are computed as energy gradients, but the loss is applied only to forces. Energy losses continue to be applied to OMol25 (and SVP via its head), anchoring the shared energy head while bypassing TZVPD energy offsets.
    \end{enumerate}
\end{itemize}

\begin{figure}[H]
  \centering
  \includegraphics[width=0.95\linewidth]{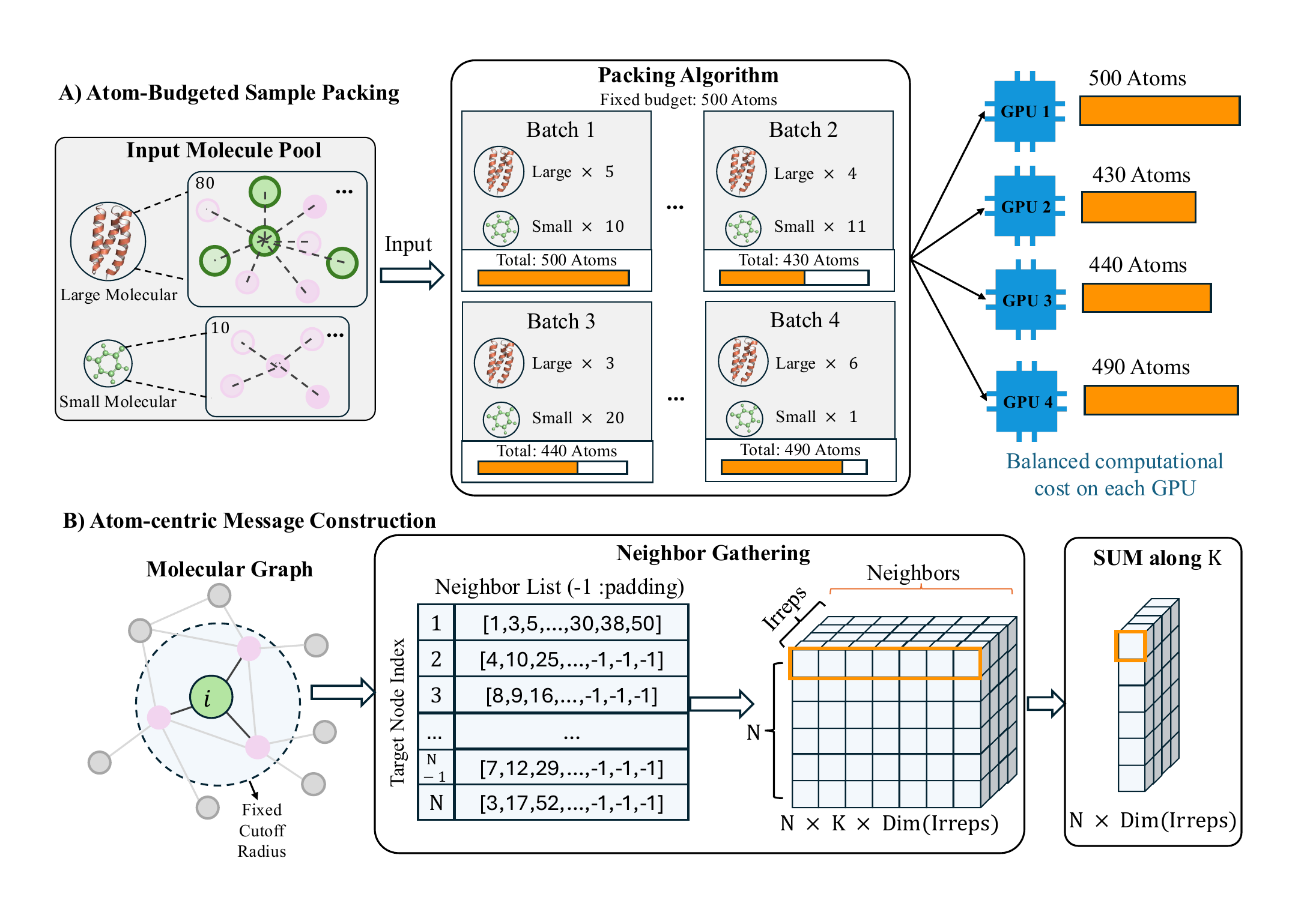}
  \caption{Atom-balanced batching and Atom-centric message passing improve efficiency by balancing GPU workload and replacing sparse edge operations with dense, regular computation.}
  \label{fig:sampling_strategy}
\end{figure}

Together, this three-stage protocol enables efficient training of a single model that is both chemically expressive and scalable to large atomistic systems. Figure~\ref{fig:sampling_strategy} summarizes the atom-balanced batching and atom-centric message construction used in our training pipeline.

\subsubsection{Atom-Balanced Data Loading Mechanism}

\textbf{Atom-Budgeted Sample Packing.}
Molecular systems exhibit highly diverse sizes.
For example, molecules in the OMol25 dataset range from 50 to 300 atoms,
while our in-house SVP and TZVPD datasets span 300--1500 atoms.
To rigorously evaluate out-of-distribution (OOD) generalization, we exclude all configurations with more than 1,200 atoms from the training set, reserving the 1,200--1,500 atom regime exclusively for the extrapolation benchmarks described in \S\ref{sec:results}.
Such variability leads to severe load imbalance when batching by the
number of molecular graphs.
To address this issue, we adopt an atom-count–based batching strategy.
Specifically, we define a maximum atom budget $\texttt{bs\_atom}$ per
batch.
After randomly shuffling the training set, molecules are greedily added
to the current batch until the accumulated atom count reaches
$\texttt{bs\_atom}$.
Remaining samples are assigned to subsequent batches.
This strategy ensures that each GPU processes a similar number of atoms,
resulting in more balanced computational cost and improved training
efficiency.

\textbf{Atom-centric Message Construction.}
Most molecular GNNs represent interactions using an edge list, where
each edge is encoded as a pair of source and target node indices.
While flexible, this edge-centric formulation requires multiple sparse
operations in CUDA, including (i) source-node gathering, (ii)
target-node gathering, and (iii) sparse aggregation.
These operations involve irregular memory access patterns and concurrent
scatter-add updates to shared target nodes, which require AtomicAdd operations
to ensure correctness and significantly limit GPU efficiency.

In molecular systems with a fixed cutoff radius, the number of neighbors
per atom is bounded by the atomic density and the cutoff volume, yielding
a small and predictable upper bound on neighborhood size.
Motivated by this observation, we adopt a node-centric neighborhood
representation.
For each target atom, we store up to $k$ neighboring source indices in a
dense tensor of shape $N \times k$, together with a binary mask indicating
valid entries.
Under this formulation, source-node features are still accessed via
indexed gather operations.
In contrast, target-node features are obtained by dense repetition rather
than sparse gathers.
Crucially, message aggregation is performed using standard dense
reductions (e.g., summation along the neighbor dimension), which
structurally eliminates concurrent write conflicts and avoids AtomicAdd operations.
By removing irregular sparse aggregation, this design
significantly improves memory access regularity and GPU utilization.

\section{Resources}
To maximize the impact of UBio-MolFM and foster reproducibility, we organize our public resources around three components: data, code/inference, and model weights.

\subsection{Data: UBio-Protein26 5M}
To facilitate community benchmarking and reproducibility, we release a curated protein-focused subset of UBio-Mol26, named \textbf{UBio-Protein26 5M}. The dataset is publicly available at:
\begin{center}
  \url{https://huggingface.co/datasets/IQuestLab/UBio-Protein26}
\end{center}

UBio-Protein26 5M is composed of protein-related configurations (solvated tripeptides and residue-centered clusters) and preserves the statistical characteristics of the full UBio-Mol26 distribution. The release contains a 5M training split (4.5M at \texttt{def2-SVP} and 0.5M at \texttt{def2-TZVPD}) and a 0.2M held-out test split, providing a standardized high-fidelity benchmark for training and evaluating macromolecular foundation models.

\begin{figure}[H]
  \centering
  \includegraphics[width=0.8\linewidth]{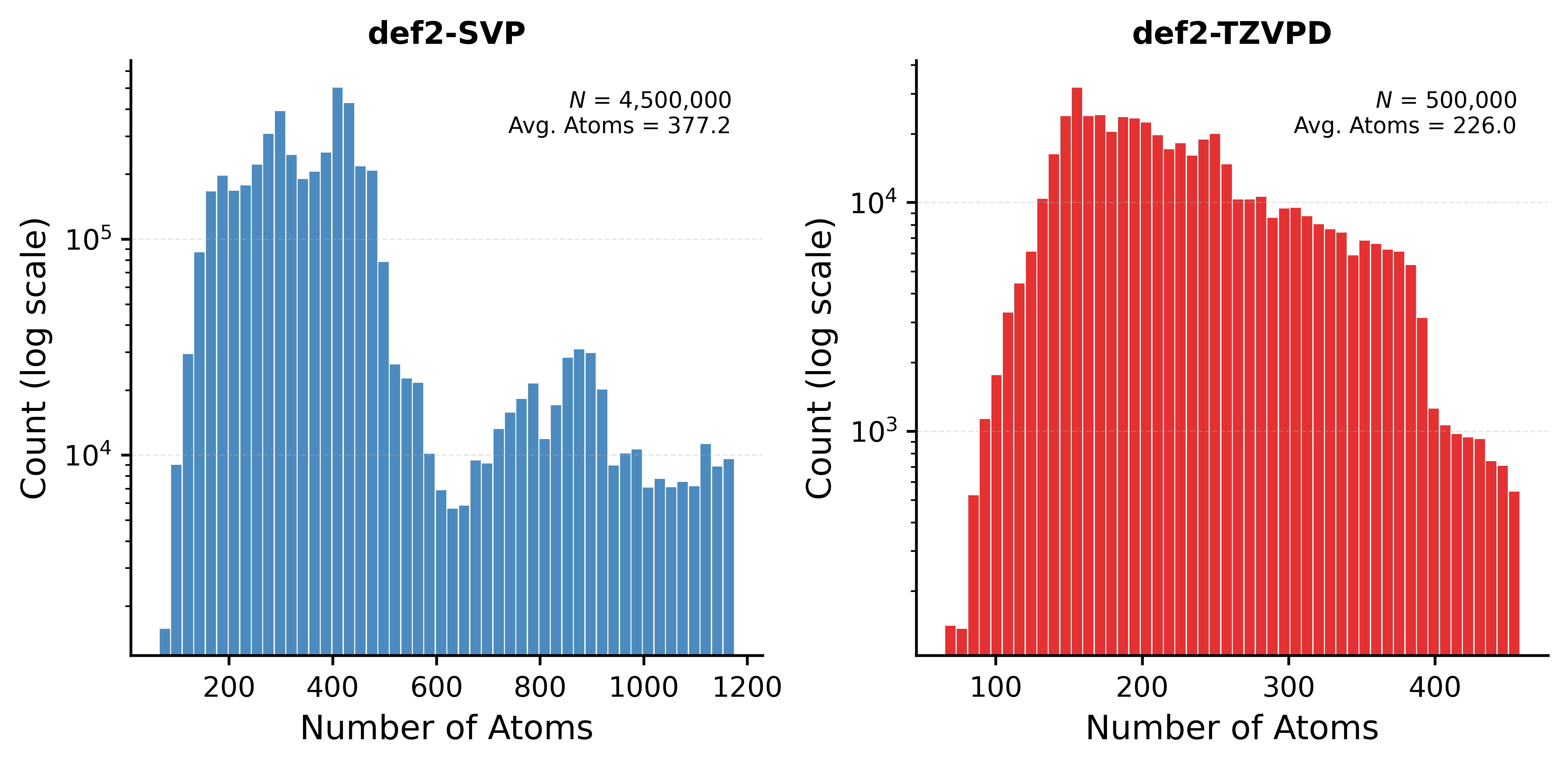}
  \caption{\textbf{Structural Distribution of UBio-Protein26 5M.} The dataset provides a comprehensive sampling of protein conformational space and local chemical environments.}
  \label{fig:protein5m_dist}
\end{figure}

The structural and chemical diversity of the training split is illustrated in Figure~\ref{fig:protein5m_dist}.

\subsection{Code and Inference}
The implementation of the \textbf{E2Former-V2} architecture is already available at:
\begin{center}
  \url{https://github.com/IQuestLab/UBio-MolFM}
\end{center}

Standardized inference scripts for UBio-MolFM, along with integrated modules for running molecular dynamics (MD) simulations, will be released concurrently with the model weights. These resources are designed for seamless deployment on commodity GPU hardware, leveraging PyTorch and Triton-optimized kernels.

\subsection{Model Weights}
The pretrained weights for the UBio-MolFM series are available on \textbf{Hugging Face} at:
\begin{center}
  \url{https://huggingface.co/IQuestLab/IQuest-UBio-MolFM-V1}
\end{center}
We provide checkpoints for Stage 3 to support biology-focused downstream applications.

\section{Related Work}
The development of UBio-MolFM builds on sustained progress in machine-learned interatomic potentials, geometric deep learning, and large-scale molecular datasets. We highlight representative work that shaped the current landscape and directly informs our design choices.

\subsection{Equivariant Neural Networks}
Equivariance to 3D rotations has become a central principle for accurate molecular modeling. Early models such as SchNet~\cite{schnet} and DimeNet~\cite{gasteiger2020directional} established strong baselines using invariant and directional descriptors. Subsequent equivariant architectures, including NequIP~\cite{nequip}, Allegro~\cite{allegro}, eSCN~\cite{escn}, and MACE~\cite{mace}, improved angular resolution via irreducible representations, enabling higher accuracy for forces and energies. Transformer-based equivariant models such as Equiformer~\cite{equiformer} and EquiformerV2~\cite{equiformerv2} further extended modeling capacity, particularly for long-range interactions. UBio-MolFM follows this trajectory but focuses on linear-scaling design choices suited for large, solvated biomolecular systems.

\subsection{Molecular Foundation Models}
The field has recently shifted toward large, general-purpose molecular foundation models. Uni-Mol~\cite{zhou2023uni} and Graphormer~\cite{ying2021do} adapted transformer architectures for molecular representation learning. DPA-2~\cite{dpa2} and MACE-Off~\cite{kovacs2023mace} advanced general-purpose interatomic potentials, while the UMA family~\cite{uma} demonstrated the benefits of scaling with massive datasets. UBio-MolFM complements these efforts by targeting macromolecular regimes, where explicit solvation, long-range electrostatics, and biological context dominate error sources.

\subsection{Data Resources and Biological Coverage}
Large, accurate datasets are a prerequisite for foundation models. SPICE and SPICE 2.0~\cite{spice, spice2} expanded public high-fidelity coverage for small molecules and solvated complexes, while OMol25~\cite{omol25} provided a major leap in scale and chemical diversity. However, these datasets remain limited in system size for macromolecular biology. UBio-Mol26 addresses this gap by integrating bottom-up enumeration and top-down sampling, in line with strategies explored in GEMS~\cite{gems}, to capture realistic biomolecular environments at larger scales.

\subsection{Scaling and Infrastructure}
Classical molecular dynamics engines such as GROMACS~\cite{gromacs}, LAMMPS~\cite{lammps}, and NAMD~\cite{namd} remain the backbone for large-scale simulation. Bridging their scale with the accuracy of ML potentials requires algorithmic efficiency and hardware-aware implementation. UBio-MolFM is designed for this regime by emphasizing linear-scaling architecture and practical inference throughput, enabling integration into large-system workflows.

\section{Contributors}
\subsection*{Research Team}
The UBio-MolFM project is a collaborative effort led by the UBio Team at IQuest Research.

\paragraph{Core Members:}
Jia Zhang (JZ)~\footnote{jialrs.z@iquestlab.com}, Lin Huang (LH), Arthur Jiang (AJ), XiaoLi Liu (XLL), Zion Wang (ZW), Jason Zhao (JAZ).

\paragraph{Project Members:}
Chu Wang (CW), HaoCheng Lu (HCL), ChengXiang Huang (CXH), JiaJun Cheng(JJC), YiYue Du (YYD).

\subsection*{Author Contributions}
\begin{itemize}
  \item \textbf{Data Generation and Analysis:} JZ, CW, HCL, AJ.
  \item \textbf{Model Design and Acceleration:} LH, JZ, ZW, CXH, JJC, YYD, XLL, JAZ, AJ.
  \item \textbf{Model Training:} LH, CXH, JZ.
  \item \textbf{Model Evaluation:} JZ, LH, CW, HCL.
  \item \textbf{Technical Report Writing:} All authors.
\end{itemize}

\bibliographystyle{unsrt}
\bibliography{ref}

\end{document}